\documentclass[10pt,aps,prc,floatfix,twocolumn,nofootinbib, superscriptaddress,preprintnumbers]{revtex4-1}
\pdfoutput=1
\usepackage{amsfonts,amsmath,amssymb,bm, xspace}
\usepackage{color,graphicx}
\usepackage{bbm}
\usepackage{dcolumn}                 
\graphicspath{{./figures/}}

\usepackage{hyperref}                

\usepackage{natbib}
\usepackage{graphicx}

\usepackage{array,physics}
\usepackage{dcolumn}
\usepackage[normalem]{ulem}
\newcolumntype{d}[1]{D{.}{.}{#1}}

\newcommand{\orcid}[1]{\href{https://orcid.org/#1}{\includegraphics[width=20pt]{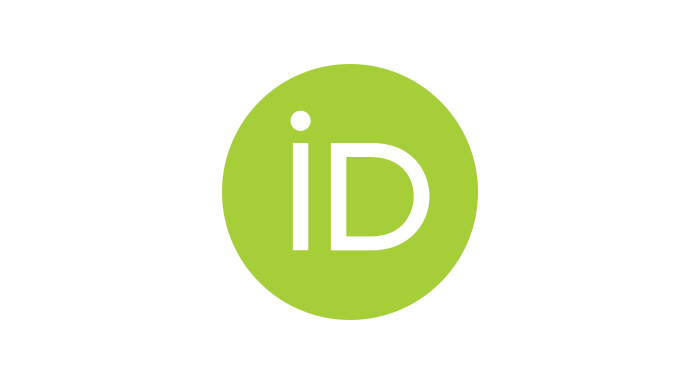}}}

\newcommand{\bat}{\Bigr\rvert}

\newcommand{\etal}{\textit{et~al.}\xspace}

\newcommand{\tov}{\mathrm{TOV}}

\newcommand{\ChiEFT}{\ensuremath{{ \rm \chi EFT}}\xspace}

\newcommand{\mm}{\mathrm{MM}}

\newcommand{\sm}{\mathrm{SM}}
\newcommand{\nm}{\mathrm{NM}}
\newcommand{\ffg}{\mathrm{FFG}}

\newcommand{\tot}{\mathrm{tot}}

\newcommand{\ex}{\mathrm{exp}}
\newcommand{\emp}{\mathrm{emp}}
\newcommand{\best}{\mathrm{best}}

\newcommand{\nuc}{\mathrm{nuc}}
\newcommand{\cl}{\mathrm{cl}}
\newcommand{\crust}{\mathrm{crust}}
\newcommand{\WS}{\mathrm{WS}}
\newcommand{\MMb}{\mathrm{MM_{bare}}}
\newcommand{\MMs}{\mathrm{MM_{m^*}}}

\newcommand{\sat}{\mathrm{sat}}
\newcommand{\sym}{\mathrm{sym}}
\newcommand{\pot}{\mathrm{pot}}
\newcommand{\bulk}{\mathrm{bulk}}
\newcommand{\fs}{\mathrm{FS}}
\newcommand{\coul}{\mathrm{Coul}}
\newcommand{\dir}{\mathrm{Dir}}
\newcommand{\exc}{\mathrm{Ex}}
\newcommand{\surf}{\mathrm{surf}}
\newcommand{\curv}{\mathrm{curv}}

\newcommand{\cc}{\mathrm{cc}}

\hypersetup{%
    pdfsubject=Paper,
    pdfkeywords={nuclear physics} {MBPT} {chiral EFT} {asymmetric nuclear matter}
    unicode=true,
    breaklinks=true,
     colorlinks   = true,
     linkcolor = blue,
     citecolor = blue,
     menucolor = blue,
     urlcolor = blue
}

\begin{document}
\title{Properties of the neutron star crust: Quantifying and correlating uncertainties with improved nuclear physics} 

\author{G. Grams\orcid{0000-0002-8635-383X}}
\affiliation{Univ Lyon, Univ Claude Bernard Lyon 1, CNRS/IN2P3, IP2I Lyon, UMR 5822, F-69622, Villeurbanne, France}

\author{R. Somasundaram\orcid{0000-0003-0427-3893}}
\affiliation{Univ Lyon, Univ Claude Bernard Lyon 1, CNRS/IN2P3, IP2I Lyon, UMR 5822, F-69622, Villeurbanne, France}
 
\author{J. Margueron\orcid{0000-0001-8743-3092}}
\affiliation{Univ Lyon, Univ Claude Bernard Lyon 1, CNRS/IN2P3, IP2I Lyon, UMR 5822, F-69622, Villeurbanne, France}

\author{S. Reddy\orcid{0000-0003-3678-6933}}
\affiliation{Institute for Nuclear Theory, University of Washington, Seattle, WA 98195-1550, USA}

\date{\today}

\begin{abstract}
A compressible liquid-drop model (CLDM) is used to correlate uncertainties associated with the properties of the neutron star (NS) crust with  theoretical estimates of the uncertainties associated with the equation of state (EOS) of homogeneous neutron and nuclear matter. For the latter, we employ recent calculations based on Hamiltonians constructed using Chiral Effective  Field theory (\ChiEFT).  Fits to experimental nuclear masses are employed to constrain the CLDM further, and we find that they disfavor some of the \ChiEFT Hamiltonians. The CLDM allows us to study the complex interplay between bulk, surface, curvature, and Coulomb contributions, and their impact on the NS crust.  It also reveals how the curvature energy alters the correlation between the surface energy and the bulk symmetry energy.\\
Our analysis quantifies how the uncertainties associated with the EOS of homogeneous matter implies significant uncertainties for the composition of the crust, its proton fraction, and the volume fraction occupied by nuclei. We find that the finite-size effects impact the crust composition, but have a negligible effect on the net isospin asymmetry of matter. The isospin asymmetry is largely determined by the bulk properties and the isospin dependence of the surface energy. The most significant uncertainties associated with matter properties in the densest regions of the crust, the precise location of the crust-core transition, are found to be strongly correlated with uncertainties associated with the Hamiltonians. By adopting a unified model to describe the crust and the core of NSs, we tighten the correlation between their global properties such as their mass-radius relationship, moment of inertia, crust thickness, and tidal deformability with uncertainties associated with the nuclear Hamiltonians. \\
\end{abstract}

\maketitle

\section{Introduction}

The understanding of neutron star (NS) properties from fundamental nuclear physics inputs requires the precise determination of the relation between nuclear physics uncertainties and dense matter predictions. This became possible recently due to advances in theoretical efforts to predict properties of nuclei and dense nuclear matter, and advances in experimental nuclear physics that are now providing more stringent constraints. In addition, recent observations of NS radii by NICER~\cite{MillerNICER2019,NICER2021}, and tidal deformabilities by the LIGO-Virgo collaboration~\cite{Abbott2017} have also reached the accuracy to sharply constrain the dense matter EOS. These developments motivate the construction of models that can provide a unified description of the EOS of the crust and the core.

Our current understanding of NS crusts suggests that it is composed of finite nuclei, usually referred to as nuclear clusters since their properties are modified by the dense matter environment and differ from those of isolated nuclei probed in the laboratories, see for instance Ref.~\cite{Haensel:2007yy}. The outer crust is dominated by the presence of an electron gas filling the whole volume in beta-equilibrium with neutrons and protons bound inside nuclear clusters. The nuclear symmetry energy controls the energy difference between neutrons and protons, and thus the isospin asymmetry inside the nuclear clusters. Electric charge neutrality is ensured by the presence of electrons, and its rapid increase with density favors the appearance of increasingly neutron-rich nuclear clusters \cite{Chamel:2008ca}. From the experimental viewpoint, neutron-rich nuclei can be produced in nuclear facilities and provide strong constraints on the properties of nuclei present in the outer crust, see Refs.~\cite{PhysRevLett.110.041101,Kreim2013} for recent updates. Nuclear clusters in the inner crust are significantly more neutron-rich and coexist with a neutron fluid and their properties cannot be directly probed by experiments. Their description relies on theoretical models that are sensitive to the properties of bulk nuclear matter and the density and isospin dependence of the nuclear surface tension \cite{Steiner:2004fi}. The neutron fluid in the inner crust is very likely to be in a superfluid state at low temperature and superfluidity is known to impact NS spin and thermal evolution

Recently, several conceptual milestones have been reached in the prediction of neutron star matter from microscopic ab initio approaches which are based on realistic nuclear Hamiltonians constrained by nucleon scattering data. In particular, Hamiltonians derived using chiral effective field theory (\ChiEFT) incorporate the symmetries of QCD and provides a systematic expansion of the operators in powers of the nucleon momenta. \ChiEFT has two distinct features: (1) it consistently include three and higher-body interactions along with the two-body interactions that are well constrained by experiments, (2) it provides a robust method to estimate errors associated with the truncation of the momentum expansion when the nucleon $p$ is small compared to the breakdown scale $\Lambda_\ChiEFT$  of the \ChiEFT. Beginning with the pioneering work of Hebeler and Schwenk \cite{Hebeler:2009iv}, several groups have used \ChiEFT to predict the EOS of homogeneous neutron-rich matter~\cite{Carbone2013,Hebeler2015,Drischler2019,Lynn2019,Rios2020}. At sufficiently low densities, neutron matter is well understood because three-body interactions are small, and the two-body neutron-neutron interaction is strongly constrained by the neutron-neutron scattering phase shifts~\cite{Heiselberg2000,Carlson2003}. However, with increasing density and correspondingly larger nucleon momenta, higher-dimension operators including three-body forces begin play an increasingly important role. For both these reasons the associated uncertainty which can be estimated quite systematically grows. The densities up to which \ChiEFT remains useful is still a matter of debate, current expectations are that it breaks-down between saturation density ($n_\sat$) and twice-$n_\sat$ \cite{Epelbaum2009,Tews:2018kmu,Drischler2020}.

Nuclear clusters in NS crust result from the equilibrium between attractive volume interaction, and repulsive surface and Coulomb interactions, at leading order. Despite recent progress in the description of finite nuclei based on chiral nuclear interaction, it is still computationally not feasible to calculate directly the properties of nuclear clusters in NS crust. Several well-motivated approximations could however be employed to predict and understand the properties of this complex system. Among them, the liquid-drop model (LDM) is a macroscopic approach that allows us to combine together the nuclear matter predictions from \ChiEFT with finite-size (FS) terms generated from a leptodermous expansion of the total energy. The compressible liquid-drop model (CLDM) includes variations of the cluster central density from one nucleus to another, through the density dependence of the bulk contribution to the total energy. The LDM and CLDM describe the collective degrees of freedom. They are different from the microscopic approach, typified by the Shell model or the energy-density functional approach, which centers around the single-particle degrees of freedom. Since the microscopic approach is the most general one, it contains the macroscopic one as an average. 

In the leptodermous expansion of the total energy~\cite{Myers1973}, the different contributions are sorted by decreasing powers of $A^{1/3}$. The order $A$ belongs to the domain of nuclear matter studies, as can be directly related to the meta-model approach for nuclear matter, the orders $A^{2/3}$ and $A^{1/3}$ characterizes the finite-size contributions, and finally, orders $A^{0}$ and below belong to the single-particle contributions, e.g. shell effects or pairing contribution, described by microscopic theories. In our study, we investigate several FS terms which are sorted by decreasing powers of $A^{1/3}$ as in the leptodermous expansion. This allows us to compare our findings with previous ones, as well as to evaluate the impact of these different terms on the NS crust properties. We actually found that the leptodermous expansion provides a good ordering of our predictions for the NS crust.

The impact of nuclear physics uncertainties on the NS crust has been analyzed in earlier work. Steiner~\cite{Steiner2008}, for instance, has constructed several NS crust using inputs from current experimental information while allowing exploration of the EOS uncertainties, in particular the one induced by the symmetry energy. Other approaches are constructed from currently available EOSs, which may not respect the low-density neutron matter expectations because they are fit to the properties of nuclei near saturation densities. In the present work, we explore both the nuclear experimental uncertainties from the knowledge of measured nuclear masses, as well as the theoretical uncertainties in the nuclear matter equation of state from many-body approach based on chiral NN and 3N interactions. This is first systematic investigation that accounts for these two sources of uncertainties. Our main findings, obtained by incorporating these two sources of information in our CLDM are:
\begin{itemize}
\item A suggested upper limit for the energy density of nuclear matter at saturation density: $\epsilon_\sat^\mathrm{max}\approx -2.30$~MeV~fm$^{-3}$. Above this limit, our model can not equilibrate the bulk and FS terms over the nuclear chart.
\item The CLDM disfavors some \ChiEFT Hamiltonians, even if the FS terms in the CLDM vary independently to the bulk term while they are correlated from first principle. This is a strong rejection since in our modeling the FS terms are optimized to fit nuclear masses, while in reality they shall be fixed. So even if -- by chance -- nature chooses to fix these FS terms to be equal to our optimization, these models would still be rejected.
\item The correlation between the surface energy (isoscalar and isovector properties) and the bulk symmetry energy depends on the considered FS model. The inclusion of the curvature contribution suggests however a typical value for the symmetry energy ($\approx 32$~MeV) where the model dependence is minimal.
\end{itemize}

We find that the present CLDM predicts the following properties for the neutron star crust: 
\begin{itemize}
\item The crust composition $(A_\cl,Z_\cl)$ is mainly determined by the considered FS model, which - for most part of them - are controlled by the experimental nuclear energies. We find however that the cluster asymmetry $I_\cl$ is much less impacted than $A_\cl$ and $Z_\cl$.
\item The crust composition $(A_\cl,Z_\cl,I_\cl)$, the proton fraction $Y_e$, and the volume fraction $u$ are essentially determined by the bulk contribution to the energy, fixing the nuclear cluster and the neutron fluid contributions. The bulk term is varied across the chiral Hamiltonians in our study.
\item In the densest region of the crust both the Hamiltonians and the surface energy isospin parameter $p_\surf$ have a dominant role in the determination of the crust-core properties and of the matter composition. These two properties are crucial for the determination of the crustal moment of inertia.
\item While not negligible, the influence of the loss function in the fit to finite nuclei as well as of the effective mass is more sub-dominant.
\end{itemize}
Finally, we analyze the global NS properties and produce uncertainties for a few observables associated to light and canonical mass NS ($1.0M_\odot$ and $1.4M_\odot$). We obtained the following results:
\begin{itemize}
\item Global properties, e.g. mass, radius, moment of inertia, tidal deformability, are critically determined by the chiral Hamiltonian properties in uniform matter. 
\item Other ingredients discussed here -- FS terms, $p_\surf$, effective mass, loss function -- play a much smaller role than the present uncertainties from astronomical observations, e.g. radius uncertainty from NICER or tidal deformability uncertainty from LIGO-Virgo gravitational wave detectors.
\item Assuming the absence of phase transition in massive NS, we found that the present uncertainty in the nucleonic effective mass modifies the spherical NS maximum mass, $M_\tov$, by about $0.15M_\odot$ at maximum.
\end{itemize}

The paper is organized as follows. Sec. \ref{sec:homogeneous} is reserved to homogeneous matter where the meta-model and the fit to $\ChiEFT$ are described. We detail the CLDM used to describe finite-size effects on the clusterized matter in Sec. \ref{sec:CLDM}. We compute and quantify uncertainties on the NS crust in Sec. \ref{sec:NS}, while details on NS macroscopic properties are discussed on Sec. \ref{sec:tov}. Finally, the conclusions of the present work are drawn on Sec. \ref{sec:conclusion}.   

\section{Homogeneous matter}
\label{sec:homogeneous}

For homogeneous matter, we consider the six Hamiltonians, H1-H5 and H7 (H6 being disregarded for not fitting well the binding energy of $^3$He), which have been generated by many-body perturbation theory (MBPT) based on chiral NN and 3N interactions~\cite{Drischler2016} and the two recent $\chi_{EFT}$ predictions from Ref.~\cite{Drischler2021}: DHS$_{L59}$ and DHS$_{L69}$. These eight predictions for nuclear matter are used to calibrate a set of eight nuclear meta-models (MM), and the version we consider here is a small extension of the original one~\cite{Margueron2018a}. This extension has already been presented in Ref.~\cite{Somasundaram2021}. The low density correction to the energy is now controlled by the function $b(\delta)=b_\sat+b_\sym \delta^2$ instead of the parameter $b$.

\begin{table*}
\centering
\tabcolsep=0.35cm
\def\arraystretch{1.6}
\begin{tabular}{lccccccc}
\hline\hline
        & \multicolumn{3}{c}{$\MMb$}& \multicolumn{3}{c}{$\MMs$}  \\
Model   & $m^*_\sat/m_N$ & $D m^*_\sat/m_N$ & $\Delta m^*_\sat/m_N$ & $m^*_\sat/m_N$ & $D m^*_\sat/m_N$  & $\Delta m^*_\sat/m_N$  \\
\hline
H1      &  1.00 & 0.00 & 0.00 & 0.59 & 0.29 & 0.43 \\
H2      & 1.00 & 0.00 & 0.00 & 0.61 & 0.26 & 0.41 \\
H3      & 1.00 & 0.00 & 0.00 & 0.61 & 0.22 & 0.34 \\
H4      & 1.00 & 0.00 & 0.00 & 0.63 & 0.25 & 0.38 \\
H5      & 1.00 & 0.00 & 0.00 & 0.66 & 0.20 & 0.33 \\
H7      &  1.00 & 0.00 & 0.00 & 0.67 & 0.26 & 0.41 \\
DHS$_{L59}$  & 1.00 & 0.00 & 0.00 & -- & -- & --\\
DHS$_{L69}$  & 1.00 & 0.00 & 0.00 & -- & -- & --\\
\hline\hline
\end{tabular}
\caption{Effective mass $m^*_\sat$ at saturation in symmetric matter, the effective mass splittings $D m^*_\sat$ and $\Delta m^*_\sat$, see Eqs.~\eqref{eq:ms}, \eqref{eq:new_splitting} and \eqref{eq:dms}. Note that $m^*_\sat$ and $D m^*_\sat$ are determined from the microscopic predictions, while $\Delta m^*_\sat$ is inferred from the MM.}
\label{table:ms}
\end{table*}

Let us briefly summarize the main ingredients of the MM.
The energy density is the sum of a kinetic and potential terms, $e_\mm=t^*+e^\pot$, where the kinetic term reads,
\begin{eqnarray}
t^{*}(n,\delta) &=& \frac{t_\sat}{2}\left(\frac{n}{n_{\textrm{sat}}}\right)^{2/3} \Big[ \frac{m}{m^*_n(n,\delta)}(1+\delta)^{5/3}\nonumber \\
&&\hspace{2cm}+\frac{m}{m^*_p(n,\delta)}(1-\delta)^{5/3} \Big] \, ,
\label{eq:tkinstar}
\end{eqnarray}
with $t_\sat=3\hbar^2 k_F^2/(10m)$ and given $\tau_3=1$($-1$) for neutrons (protons), the effective mass reads
\begin{eqnarray}
\frac{m}{m^{*}_{\tau}(n,\delta)} &=& 1 + \left(\frac{\kappa_{\textrm{sat}}}{n_{\textrm{sat}}}+\tau_3\delta\frac{\kappa_{\textrm{sym}}}{n_{\textrm{sat}}}\right) n .
\label{eq:ms}
\end{eqnarray}

In the present work we explore the effect of including or not the effective mass on the kinetic energy. We show results with the effective mass equal to the bare mass $m^*_{\tau} = m_{\tau}$, and also with the effective mass given by Eq. \eqref{eq:ms}. In order to obtain the parameters $\kappa_{\textrm{sat}}$ and $\kappa_{\textrm{sym}}$ for the Hamiltonians we consider here, we first derive the density dependent Landau effective mass for neutrons from its single-particle spectrum, as it was done in Ref.~\cite{Somasundaram2021}. Then Eq.~\eqref{eq:ms}, with $\tau = n$, was fit to this quantity in symmetric matter (SM) and neutron matter (NM) for each individual Hamiltonian. As far as the finite residuals of the fits are concerned, the uncertainties on the parameters $\kappa_{\textrm{sat}}$ and $\kappa_{\textrm{sym}}$ are negligible for all Hamiltonians. There is however a spread in the results of the fit across different Hamiltonians reflecting an intrinsic uncertainty in the $\chi$EFT predictions. Ref.~\cite{Somasundaram2021} performed a Bayesian quantification of this uncertainty, whereas in this work we probe the $\chi$EFT uncertainty by making distinct predictions for each individual Hamiltonian and then monitoring the dispersion among these predictions. 

The isospin splitting of the effective mass is defined as,
\begin{eqnarray}
\Delta m_{\sat}^*&=&m^*_n(n=n_{\sat},\delta=1)-m^*_p(n=n_{\sat},\delta=1) \nonumber \\
&=& \frac{-2\kappa_{\sym}}{(1+\kappa_{\sat})^2-\kappa_{\sym}^2} m_N \, ,
\label{eq:dms}
\end{eqnarray}
where we fixed $m_n=m_p=m_N$. Sometimes, the proton effective mass in NM is not calculated from microscopic approaches, but SM and NM calculations are performed. In these cases, it is interesting to consider the difference between the neutron effective masses in NM and SM defined as,
\begin{equation}
D m_{\sat}^*=m^*_n(n=n_{\sat},\delta=1)-m^*_n(n=n_{\sat},\delta=0).
\label{eq:new_splitting}
\end{equation}

The effective mass, the splitting of the effective mass $\Delta m_{\sat}^*$ and $Dm_{\sat}^*$ for the eight Hamiltonians are shown in Table~\ref{table:ms}. We find that $Dm_{\sat}^*$, which is obtained directly from the MBPT single-particle spectrum as explained before, is in good agreement with our previous study in Ref.~\cite{Somasundaram2021}. The splitting of the effective mass can be inferred using the MM parametrization of the effective mass, see Eq.~\eqref{eq:ms}. We found a very good correlation between $Dm_{\sat}^*$ and $\Delta m_{\sat}^*$, suggesting that $\Delta m_{\sat}^*$ is about 50\% larger than $D m_{\sat}^*$. Finally, for DHS$_{L59}$ and DHS$_{L69}$ we only consider the bare mass case.

In the MM, the potential term is expressed as,
\begin{eqnarray}
e^\pot(n,\delta) &=& \sum_{j=0}^{N} \frac{1}{j!}\left(v_{\sat,j}+v_{\sym 2,j}\delta^2\right) x^j \nonumber \\
&& + e^{\pot, \mathrm{low-n}}
\end{eqnarray}
with the low-density correction expressed as
\begin{eqnarray}
e^{\pot, \mathrm{low-n}} = v^\mathrm{low-n}(\delta) \,x^{N+1} e^{-b(\delta) n / n_\sat^\emp} \label{eq:lown}\, , 
\end{eqnarray}
with $v^\mathrm{low-n}(\delta)=v_\sat^\mathrm{low-n}+v_\sym^\mathrm{low-n}\delta^2$. The parameters $v_\sat^\mathrm{low-n}$ and $v_\sym^\mathrm{low-n}$ are fixed by the condition that $e^\pot(n=0,\delta)=0$ for $\delta=0$ and $\delta=1$, and the parameters $b_\sat$ and $b_\sym$ are adjusted to reproduce the very low density dependence of the MBPT calculations, see Ref.~\cite{Somasundaram2021} for more details. With the MM, the binding energy $e_\mm(n,\delta)$ can be obtained for any arbitrary value of the density $n$ and the isospin asymmetry parameter $\delta$. There are two interesting limits which are SM and NM defined as $e_\sm(n)=e_\mm(n,\delta=0)$ and $e_\nm(n)=e_\mm(n,\delta=1)$.

Table \ref{table:xhi2chiEFT} shows the residual $\chi^2$ values for the fit, where the $\chi^2$ is defined as 
\begin{equation}
    \chi^2 = \frac{1}{2} \sum_i \bigg( \frac{e_{\textrm{data},i}-e_{\textrm{MM},i}}{\sigma_i} \bigg)^2 ,
    \label{eq:chi2}
\end{equation}
where $\sigma_i$ was taken to be a $10\%$ uncertainty on the data. The fit was performed by minimizing the $\chi^2$ using the standard Levenberg-Marquardt algorithm implemented in python's scipy package. The impact of the effective mass is very small. At most it improves the reduced $\chi^2$ by 10\% in SM, no effect in NM. In addition, in the density region out of the data, the impact of the effective mass is also very small. This will also be confirmed in Sec.~\ref{sec:tov} where the impact of the effective mass on the mass-radius relation will be presented. In Table~\ref{table:xhi2chiEFT}, for Hamiltonians H1-H5 and H7, the number of data points N is 11 (11) for SM (NM) when $0.4<k_{Fn}<1.0$ and 22 (24) for SM (NM) when $0.4<k_{Fn}<1.6$. For the other two Hamiltonians, N is 17 (9) in SM (NM) when $0.4<k_{Fn}<1.6$ and 2 (-) for SM (NM) when $0.4<k_{Fn}<1.0$.

\begin{table}[t]
\centering
\tabcolsep=0.18cm
\def\arraystretch{1.6}
\begin{tabular}{lccccc}
\hline\hline  
$\chi^2/N$ & \multicolumn{2}{c}{$0.4<k_{Fn}<1.6$}& \multicolumn{2}{c}{$0.4<k_{Fn}<1.0$}\\
Model  & SM  & NM  & SM    & NM     \\
\hline
H1      & 0.56(0.48) & 0.03(0.02) & 0.60(0.52) & 0.02(0.02) \\
H2      & 0.55(0.48) & 0.03(0.02)  & 0.58(0.51) & 0.02(0.02) \\
H3      & 0.35(0.30) & 0.01(0.01)  & 0.46(0.41) & 0.01(0.01) \\
H4      & 0.55(0.50) & 0.03(0.03) & 0.57(0.51) & 0.02(0.02) \\
H5      & 0.55(0.51) & 0.03(0.03) & 0.53(0.49) & 0.02(0.04) \\
H7      & 0.16(0.14) & 0.05(0.05)  & 0.27(0.23) & 0.04(0.07)\\
DHS$_{L59}$  & 0.76 & 0.01 & 1.6 & -- \\
DHS$_{L69}$  & 1.67 & 0.09  & 3.51 & -- \\
\hline\hline
\end{tabular}
\caption{Reduced $\chi^2/N$ in SM and NM considering the bare (effective) nucleon mass reflecting the residuals between the MBPT data and the MM, for $0.4<k_{Fn}<1.6$ on left, and $0.4<k_{Fn}<1.0$~fm$^{-1}$ on the two right columns. All quoted values are dimensionless, see Eq.~\eqref{eq:chi2}.  }
\label{table:xhi2chiEFT}
\end{table}

\begin{figure*}[t]
\centering
\includegraphics[scale=0.42]{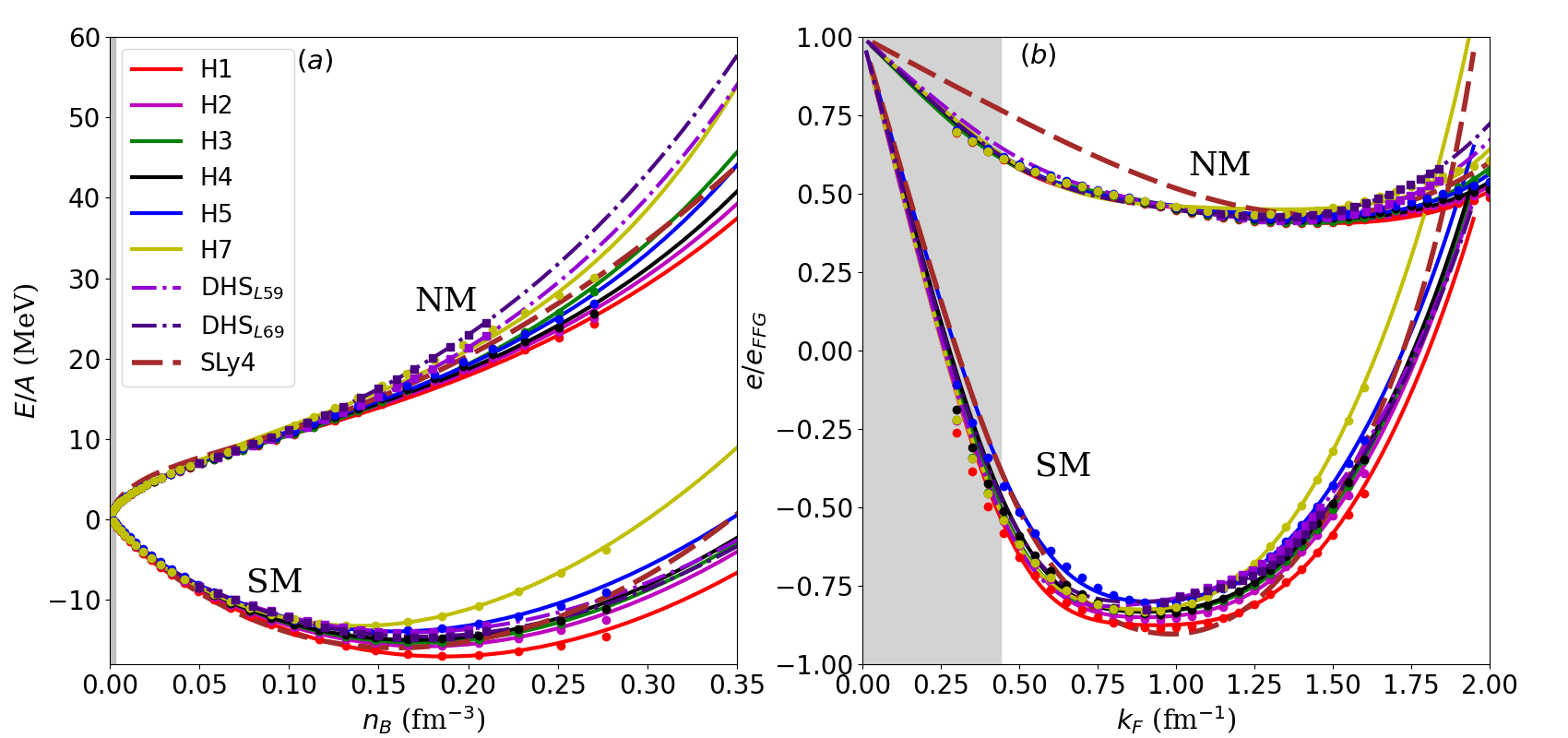}\\
\includegraphics[scale=0.42]{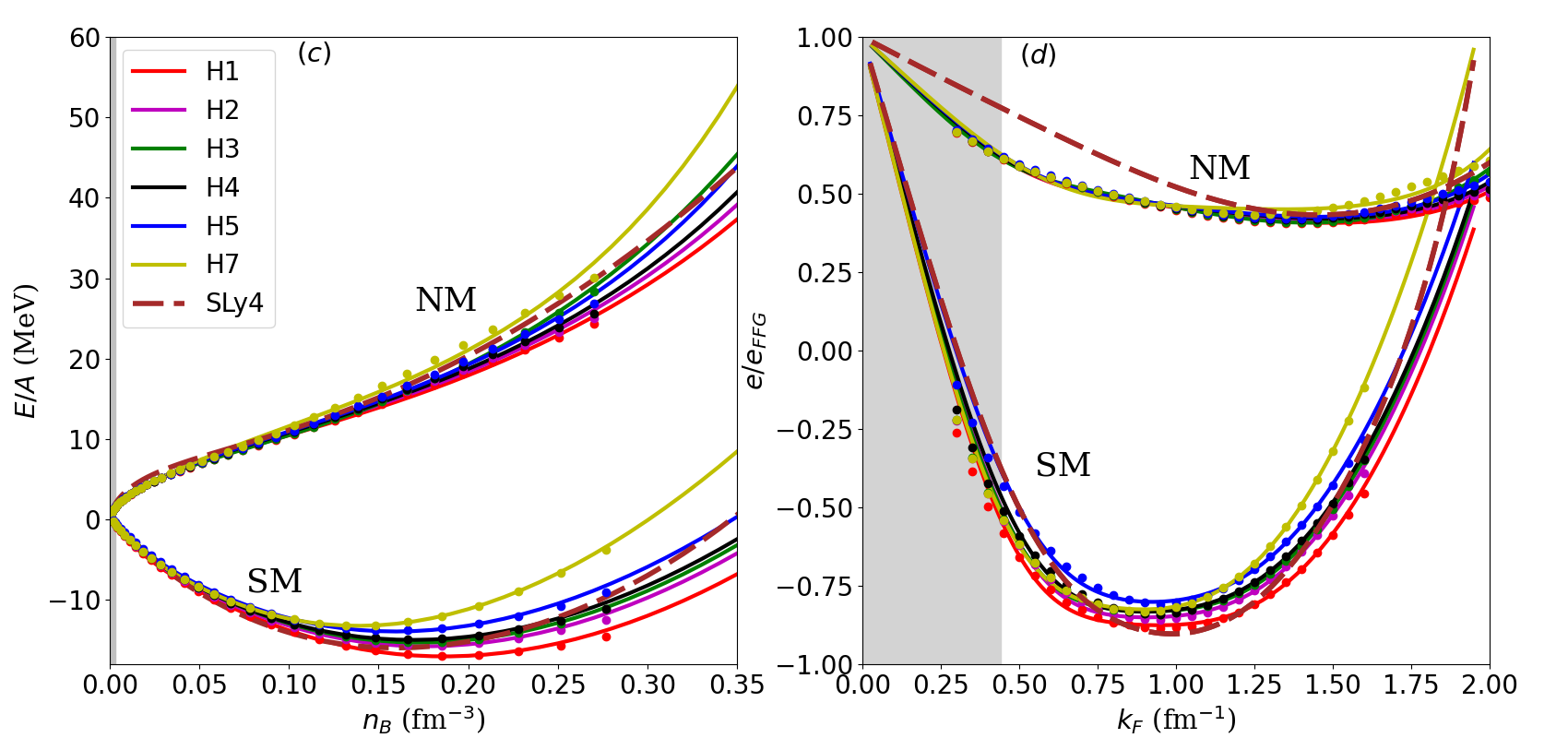}
\caption{Left:  Energy per particle as function of the baryon density, in NM and SM. Right: Energy per particle normalized to the free Fermi gas energy ($E_\ffg=(3/5)E_F$ with $E_F=\hbar^2k_F^2/(2m)$ and $k_F$ the Fermi momentum) in SM and NM function of the neutron Fermi momentum $k_{F_n}$ for the six Hamiltonians H1-H7 (except H6). Top panels include DHS Hamiltonians for comparison. Data from the original model are plotted in dots (squares) for H1-H5 and H7  (DHS$_{L59}$ and DHS$_{L69}$), with the same color as the MM version. Bottom panels include the nucleon effective mass. }
\label{fig:SMNM}
\end{figure*}

A detail comparison of the MBPT calculations (dots) and the MM fit (lines) is shown in Fig.~\ref{fig:SMNM}, the well known SLy4 Skyrme model prediction is also shown for reference. For the bare mass case (top panels) we also represent the recent MBPT calculations DHS$_{L59}$ and DHS$_{L69}$. The three models DHS$_{L59}$, DHS$_{L69}$ and H7 present stiffer NM energy at high densities compared to H1-H5. There is a very good agreement between the data and the MM down to $k_F=0.4$~fm$^{-1}$ (in density $n \approx 0.004$~fm$^{-3}$), as shown in panels (b) and (d) for instance. The vertical gray band shows the region where the fit is less accurate. In the following, the MM calibrated on the Hamiltonian H$n$ will be labelled as H$n_\mm$.

\begin{table*}[t]
\centering
\tabcolsep=0.08cm
\def\arraystretch{1.6}
\begin{tabular}{lcccccccccccc}
\hline\hline
Model & $E_\sat$ & $n_\sat$ & $\epsilon_\sat$ & $K_\sat$ & $E_\sym$ &  $L_\sym$ & $K_\sym$ & $E_{\sym,2}$ &  $L_{\sym,2}$ &  $K_{\sym,2}$ & $b_\sat$ & $b_\sym$ \\
   &(MeV) & (fm$^{-3}$) &  (MeV) & (MeV) & (MeV) & (MeV) & (MeV) & (MeV) & (MeV) & \\
\hline\hline
H1$_\mm$&  -17.0 & 0.186 & -3.17 &  261 & 33.8 & 46.8 & -154 & 33.0/31.4 & 45.3/37.2 & -152/-169  &  10.73/11.37 & 9.91/10.46\\
H2$_\mm$ &  -15.8 & 0.176 & -2.78 & 237 & 32.0 & 43.9 & -144 & 31.3/29.9 & 42.4/35.4 & -142/-156 & 9.14/9.59 & 8.86/9.34 \\
H3$_\mm$ &  -15.3 & 0.173 & -2.65 & 232 & 31.8 & 50.6 & -96 & 31.0/29.8 & 49.1/42.8 & -94/-108 & 9.83/10.35 & 18.17/20.56 \\
H4$_\mm$&  -15.0 & 0.169 & -2.54 & 223 & 31.0 & 42.1 & -138 & 30.2/29.0 & 40.7/34.5 & -136/-148 & 8.03/8.37 & 8.23/8.70  \\
H5$_\mm$&  -13.9 & 0.159 & -2.22 & 207 & 29.4 & 40.2 & -128 & 28.7/27.7 & 38.8/33.9 & -127/-137 & 6.22/6.41 & 7.70/8.24 \\
H7$_\mm$ &  -13.2 & 0.139 & -1.84 & 201 & 28.1 & 36.5 & -150 & 27.4/26.4 & 35.3/30.3 & -148/-158 & 8.98/9.40 & -1.12/-1.46\\
DHS$^{L59}_\mm$ &  -14.0 & 0.168 & -2.36 & 200 & 31.4 & 58.9 & -30 & 30.6 & 57.4 & -28 & 9.00 & 10.00\\
DHS$^{L69}_\mm$ &  -14.6 & 0.173 & -2.53 & 216 & 33.7 & 69.0 & -20 & 33.0 & 67.5 & -19 & 9.00 & 10.00\\
\hline
SLy4$_\MMs$&  -16.0 &   0.160 & -2.55 &  230 &    32.0 &   46.0 & -120 &    31.3 &   44.7 & -118 &     6.90 & 0 \\
\hline
Exp.~\cite{Margueron2018a} &  -15.8(3) & 0.155(5) & -2.45(12) & 230(30) & -- & -- & -- & 32(2) & 50(10) & -100(100)  & -- & -- \\
\hline\hline
\end{tabular}
\caption{Empirical parameters for the Hamiltonians derived from chiral EFT used in the present work. The energy density at saturation density is defined as $\epsilon_\sat=n_\sat E_\sat$. The last two columns show the low density correction parameters $b_{\sat} $ and $b_{\sym}$. For the empirical parameters for which the effective mass plays a role, we give the value obtained with the bare mass first and then the one obtained with the effective mass. }
\label{tab:empirical:Hamiltonians}
\end{table*}

Note in Fig. \ref{fig:SMNM}(b) that
the MM reproduces very well the NM energies as predicted by $\ChiEFT$, at variance with SLy4 which overestimates the energy per particle at low density. This is indeed a general feature of Skyrme interactions. We recently analyzed the impact of this systematical differences between $\ChiEFT$ and Skyrme SLy4 in low density NM on the crust EOS within the CLDM~\cite{Grams:2021}. We found that some observables are very sensitive to these differences, e.g. energy density, pressure, sound speed, while other are much less impacted, e.g. cluster configuration ($A_\cl$, $Z_\cl$, $I_\cl$), which are mostly determined by experimental nuclear masses. Having a good description of NM as predicted by $\ChiEFT$ is however important to predict NS crust properties.

The symmetry energy is defined as the energy difference between NM and SM,
\begin{equation}
e_\sym(n) = e_\nm(n) - e_\sm(n) \, ,
\label{eq:esym}
\end{equation}
and the quadratic contribution to the symmetry energy reads,
\begin{equation}
e_{\sym,2}(n) = \frac 1 2 \frac{\partial^2 e(n,\delta)}{\partial \delta^2}\bat_{\delta=0} \, ,
\label{eq:esym2}
\end{equation}

The topological properties of the energy per particle around saturation density, with empirical expectation $n_\sat^\emp \approx 0.155(5)$~fm$^{-3}$, are encoded into the nuclear empirical parameters (NEP), e.g. $E_\sat$, $E_\sym$, $E_{\sym,2}$, which are defined as,
\begin{eqnarray}
e_\sm(n) &=& E_\sat + \frac 1 2 K_\sat x^2 + \frac{1}{6} Q_\sat x^3 
+ \frac{1}{24} Z_\sat x^4+\dots \, , \nonumber \\ \\
e_\sym(n) &=& E_\sym + L_\sym x + \frac 1 2 K_\sym x^2 + \frac{1}{6} Q_\sym x^3 \nonumber \\ 
&&\hspace{1cm}+ \frac{1}{24} Z_\sym x^4+\dots \, , \\
e_{\sym,2}(n) &=& E_{\sym,2} + L_{\sym,2} x + \frac 1 2 K_{\sym,2} x^2 + \frac{1}{6} Q_{\sym,2} x^3 \nonumber \\ 
&&\hspace{1cm}+ \frac{1}{24} Z_{\sym,2} x^4+\dots \, , 
\end{eqnarray}
where the density expansion parameter is defined as $x=(n-n_\sat)/(3n_\sat)$.

The low order NEP of the 8 Hamiltonians are given in Tab.~\ref{tab:empirical:Hamiltonians}. The values are rounded and the uncertainties in these parameters are smaller than the rounding. These uncertainties are thus not given. The dispersion of the NEPs across the different Hamiltonians however serves to capture uncertainties intrinsic to the $\chi$EFT expansion. The saturation energy and density of H2$_\mm$-H4$_\mm$ agree reasonably well with the empirical ones determined from experimental data, see for instance Ref.~\cite{Margueron2018a}, while the saturation energy of H5$_\mm$ and H7$_\mm$ is higher than the expected one. The saturation density of H7$_\mm$ is also quite lower than the empirical one. For H1$_\mm$ the saturation density is higher than the expected one, and the saturation energy is lower. In the following, we will confirm that the fit to the experimental masses is poorer for H5$_\mm$ and H7$_\mm$ compared to the other Hamiltonians, as we can already anticipate from their empirical properties.

It is well known that the high order empirical parameters $Q_{\sat/\sym}$ and $Z_{\sat/\sym}$ are not constrained by $\chi_{EFT}$ calculations in uniform matter, since $\chi_{EFT}$ is limited to low densities, $n<2n_\sat$~\cite{Somasundaram2021}. In the present study, we fix the following values for all Hamiltonians: $Q_\sat = -220$~MeV, $Z_\sat = -200$~MeV, $Q_\sym = 700$~MeV, and $Z_\sym = 500$~MeV. This choice allows all the considered Hamiltonians to reach the astrophysical constraint related to the observed maximum mass of NSs, which is 2.0(1) $M_{\odot}$~\cite{Demorest2010,Antoniadis2013}. The recent measurement for PSR J0740+6620, suggesting $M_\tov\geq 2.14(10)$ M$_\odot$ is also compatible with this constraint~\cite{Cromartie2020}.

\begin{figure*}[t]
\centering
\includegraphics[scale=0.43]{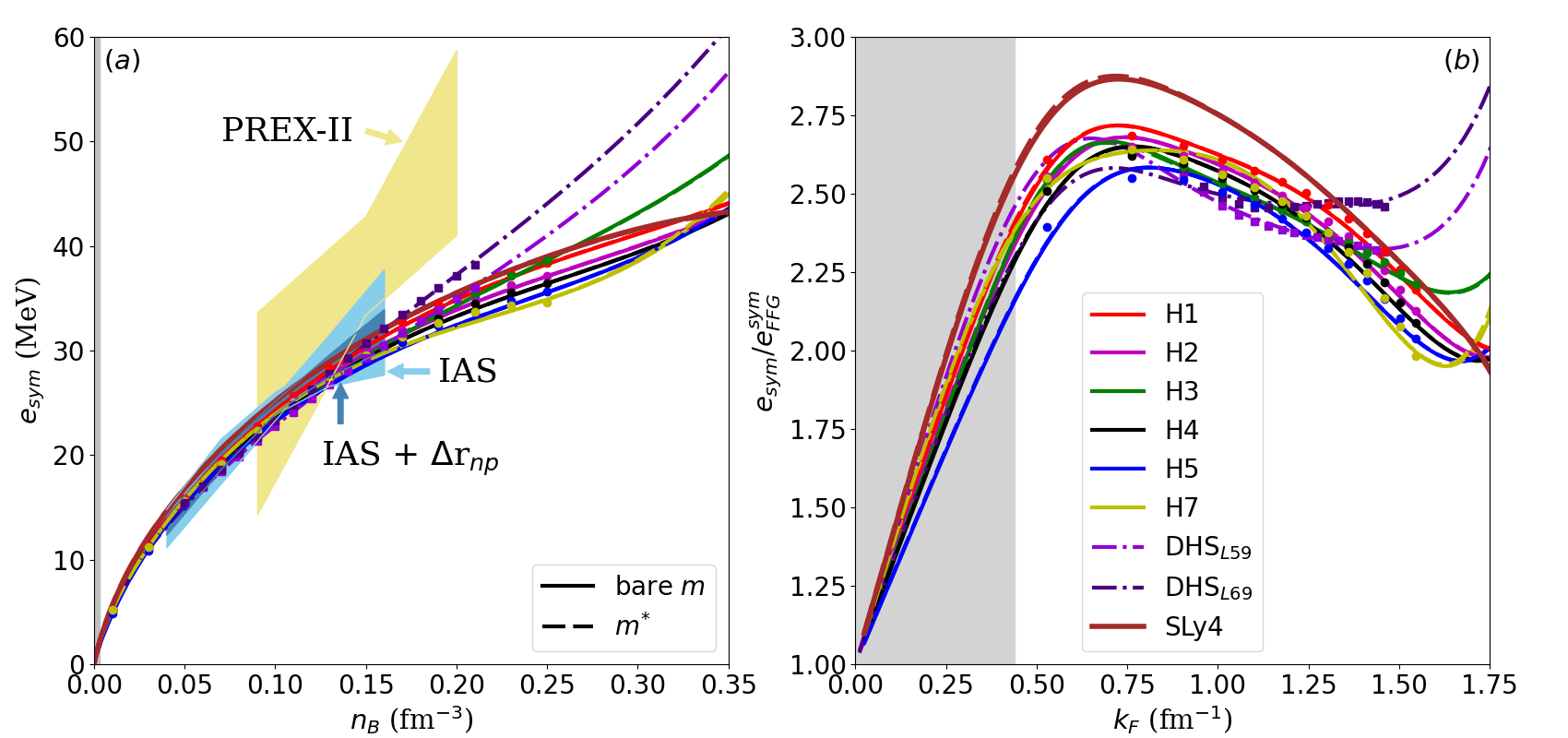}
\caption{Left: Symmetry energy w.r.t the baryon density for H1-H5 and H7, DHS$_{L59}$, DHS$_{L69}$ and SLy4. Yellow band show constrain from neutron skin experiments by PREX-II. Blue bands show constrains from isobaric analog state IAS (IAS + neutron skin, $\Delta r_{np}$). Right: symmetry energy normalized to the free Fermi gas symmetry energy ($E_{\sym,\ffg}=(3/5)(E_{F,\nm}-E_{F,\sm})$) w.r.t the  Fermi momentum $k_{F}$. Continuous (dashed) lines considers bare (effective) nucleon mass.}
\label{fig:ESYM}
\end{figure*}

The symmetry energy $e_\sym(n)$ is shown in Fig.~\ref{fig:ESYM}.
Left panel shows in light (dark) blue band the constrains from isobaric analog state IAS (IAS + neutron skin, $\Delta r_{np}$), from Ref.~\cite{Danielewicz2014}. In yellow are shown the PREX-II predictions for the symmetry energy, where we vary $E_\sym = 38.1 \pm 4.7$ MeV and $L_\sym = 106 \pm 37$ MeV as suggested by the publication~\cite{Brendan2021}. The symbols represent the MBPT calculations~\cite{Drischler2016,Drischler2021} as in Fig. \ref{fig:SMNM}. Right panel shows the symmetry energy, $e_{\sym}$, normalized by the free fermi gas symmetry energy. Note the higher value for the symmetry energy predicted by SLy4 compared to the Hamiltonians at low densities. This is a direct effect of the high NM energy predicted by SLy4, as seen in Fig. \ref{fig:SMNM}(b) and (d). The solid lines shows the MM with the bare mass while the dashed lines include the corrections induced by the effective mass, see Eqs.~\eqref{eq:tkinstar} and \eqref{eq:ms}. Note that the effect of the effective mass is very small on these curves.

\begin{figure}[t]
\centering
\includegraphics[scale=0.30]{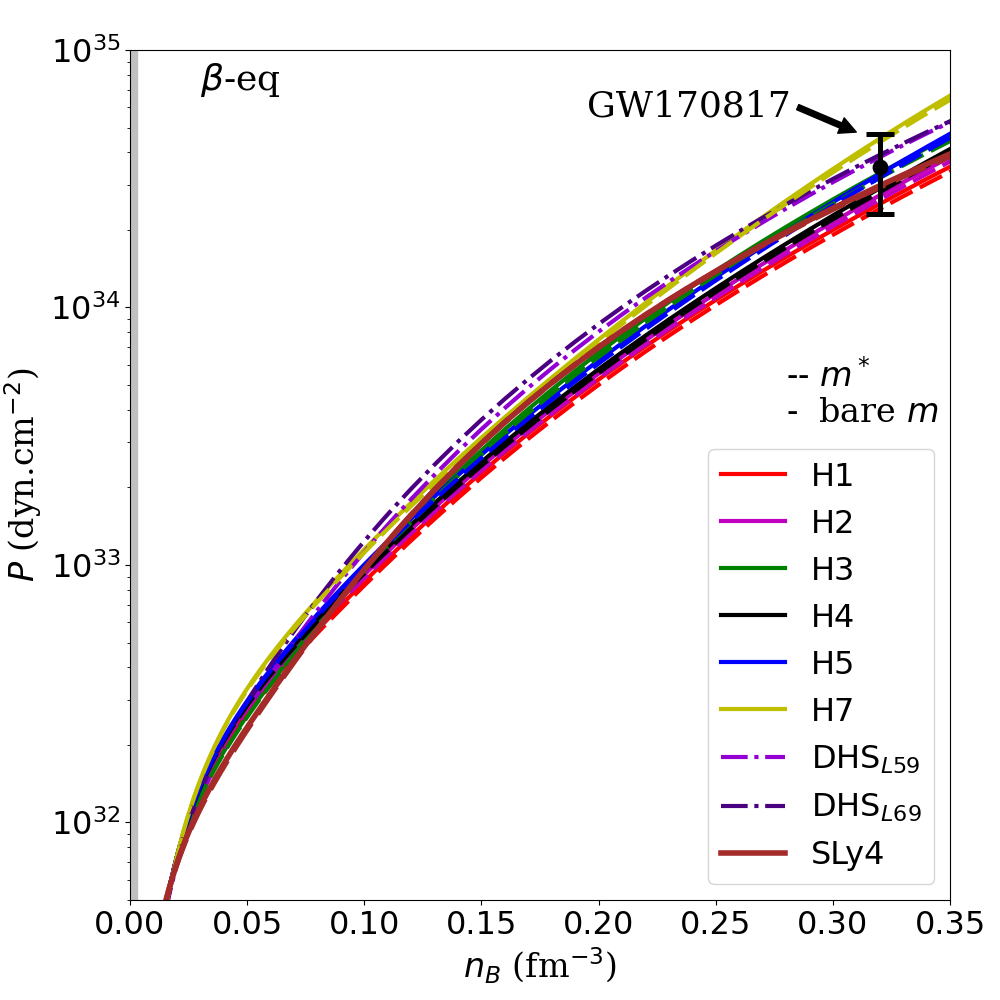}
\caption{Pressure as function of the baryon density in $\beta$-equilibrium for the six Hamiltonians, DHS$_{L59}$, DHS$_{L69}$ and SLy4. Continuous (dashed) lines considers bare (effective) nucleon mass.  Error bar shows the constrain for the pressure at 2$ n_{\sat}$ inferred by the LIGO/Virgo collaboration with the GW170817 observation at 90$\%$ credible level. }
\label{fig:pressBetaUnif}
\end{figure}

There is a disagreement between  $\chi_{EFT}$ predictions and the recent constrain from PREX-II~\cite{Brendan2021} for the symmetry energy and its density dependence, as shown in Fig. \ref{fig:ESYM}. To reproduce PREX-II predictions not only a large value for $L_\sym $ is necessary, like $DHS_{L59}$ and $DHS_{L69}$, but also for $E_\sym $.

The $\beta$-equilibrium with e and $\mu$ satisfies the following equations,
\begin{equation}
\mu_n = \mu_p + \mu_e \, , \quad \mu_e = \mu_\mu \, .
\end{equation}
The total pressure, including baryon and lepton contributions, at $\beta$-equilibrium is shown in Fig.~\ref{fig:pressBetaUnif} for the eight Hamiltonians and SLy4. The prediction by the LIGO-Virgo collaboration inferred from GW170817 for the pressure at 2 $n_\sat$~\cite{Abbott2018} is also shown.
Fig. \ref{fig:pressBetaUnif} shows that the eight models are in well agreement with GW170817 constrain. The impact of the effective mass $m^*$ (dashed lines versus solid lines with the bare mass) remains small for the pressure. 

In summary, we have shown that the MM can accurately reproduce the MBPT predictions for each of the considered eight Hamiltonians. In addition, we impose a prescription for the high order empirical parameters that allows the present extension of the MBPT to reach the astrophysical constraint for the TOV mass. We have also checked that all the present Hamiltonians are consistent with the inferred pressure at 2 $n_\sat$ by the LIGO-Virgo collaboration.

\section{The compressible liquid-drop model for finite nuclear systems}
\label{sec:CLDM}

Finite nuclear systems results from the equilibrium between the bulk attraction, as seen in homogeneous matter, and the surface repulsion, which originates mostly from the surface tension and the Coulomb repulsion. In the present study, we consider various extensions of the compressible liquid-drop model (CLDM), see for instance the seminal BBP model~\cite{Baym1971}, which can be justified from the leptodermous expansion~\cite{Myers1973}.

In the crust of NS, we consider the following composition: the nuclear clusters are composed of neutrons and protons, which are described by $A_\cl$ and $I_\cl$, being the mass number and the isospin asymmetry. The neutron and proton particle numbers in the nuclear clusters are obtained from: $N_\cl=A_\cl(1+I_\cl)/2$ and $Z_\cl=A_\cl(1-I_\cl)/2$. In addition, clusters are embedded in electron and neutron gases, described by their uniformly distributed densities $n_e$ and $n_{ng}$. We have implicitly assumed the  r-representation~\cite{Papakonstantinou2013} for the Wigner-Seitz cell. In this representation, the particles in the cluster volume $V_\cl$ are in equilibrium with the ones in the outside volume $V_\WS-V_\cl$, where $V_\WS$ is the Wigner-Seitz volume. There are therefore five variables in total (four particles and one volume), but the actual variables can be any combination of these variables. In the present study, we consider the following ones: $A_\cl$, $I_\cl$, $n_\cl$, $n_e$ and $n_{ng}$, where $n_\cl=A_\cl/V_\cl$ as in Ref.~\cite{Carreau2019a}.

The total cluster energy in NS crust is expressed as the sum of the independent contributions from the clusters $E_\cl$, the electrons $E_e$ and the neutron gas $E_{ng}$,
\begin{eqnarray}
E_\tot(A_\cl,I_\cl,n_\cl,n_e,n_{ng}) &=& E_\cl(A_\cl,I_\cl,n_\cl) + E_e(n_e) \nonumber \\
&& + E_{ng}(n_{ng}) \, .
\label{eq:etot}
\end{eqnarray}
The cluster contribution in the CLDM is expressed as a bulk energy contribution, determined from homogeneous matter, and a finite-size contribution, including Coulomb, surface, curvature terms at leading orders. 
The cluster binding energy contributing to Eq.~\eqref{eq:etot} is given by,
\begin{equation}
E_\cl(A_\cl,I_\cl,n_\cl,n_e) = E_\bulk(I_\cl,n_\cl) + E_\fs(A_\cl,I_\cl,n_\cl,n_e) \, ,
\label{eq:ecl}
\end{equation}
$n_\cl$ being the cluster central density. In the CLDM, the clusters are considered with a uniform density, and the neutron and proton radii are identically equal to the cluster radius $R_\cl$ (there is no neutron skin in the present model). There is no smooth decrease of the density profile near the cluster radius, at variance with the droplet model~\cite{Myers1980} or Thomas-Fermi approaches~\cite{Vinas2017,Haensel2001}.

In the NS crust, the total density $n_B=(A_\cl+N_g)/V_\WS$ is further imposed, contributing to fix one constraint among the five independent variables. This constraint is treated with the Lagrange multiplier technique, as suggested in Ref.~\cite{Gulminelli2015} for instance.

The energy of an isolated nucleus such as the ones present on Earth is simply defined as,
\begin{eqnarray}
E_\nuc(A_\cl,I_\cl,n_\cl) &=& E_\bulk(A_\cl,I_\cl,n_\cl) \nonumber \\
&& + E_\fs(A_\cl,I_\cl,n_\cl,n_e=0) \, .
\label{eq:enuc}
\end{eqnarray}
Note that there are only three independent variables in this case. There are no electron and neutron gases surrounding the nuclear cluster. In addition, note that the Wigner-Seitz volume $V_\WS$ is undefined (it is indeed infinite for isolated nuclei) but the cluster volume $V_\cl$ is.

In the present CLDM, the global asymmetry of the cluster $I_\cl=(N_\cl-Z_\cl)/A_\cl$ coincides with the cluster bulk asymmetry $\delta_\cl=(n_{\cl,n}-n_{\cl,p})/n_B$ since the neutron or proton skin are not considered here. Note however that neutron skin has been considered in Ref.~\cite{Steiner2008} by introducing a fit parameter $\zeta$ relating $I_\cl$ and $\delta_\cl$, as $\delta_\cl=\zeta I_\cl$. If $\zeta$ is unity, there is no skin, while if $\zeta<1$, then all nuclei with $N_\cl > Z_\cl$ will have a neutron skin ($R_n>R_p$). In reality, the parameter $\zeta$ is function of $A_\cl$ and $Z_\cl$, see Ref.~\cite{Myers1969}, as well as of the nuclear interaction, as illustrated by the correlation between the neutron skin in $^{208}$Pb and the slope of the symmetry energy $L_\sym$~\cite{Brown2000,RocaMaza2011}. It is thus a strong approximation to impose the relation $\delta_\cl=\zeta I_\cl$ with $\zeta$ constant that we prefer not to consider here. The skin contribution modifies the Coulomb term by using the proton radius instead of the cluster radius, which modified the Coulomb energy by a factor proportional to the difference between the bulk asymmetry $\delta_\cl$ and the global asymmetry $I_\cl$. This term increases with the nuclear size and asymmetry and can be important for nuclear clusters present close to the crust-core transition. However, this modification is small compared to the leading order terms considered here and we follow the procedure of recent works  \cite{Carreau2019a, Carreau2019b} and neglect the presence of neutron skin in the present study. In a future development, a consistent derivation in the spirit of Ref.~\cite{Myers1969} for instance will be considered.

The CLDM we consider is comparable to the pioneering BBP model~\cite{Baym1971} and is well suited to analyze the origin of the uncertainties in the predictions of the NS crust. More microscopic models for the crust have indeed been developed, see for instance Ref.~\cite{Negele1973} and recent efforts in Refs.~\cite{Sharma2015}. While being less accurate than microscopic models in reproducing finite nuclei, the present CLDM allows us a better understanding of the various features influencing the properties of the NS crust, which are difficult to analyze in a microscopic model. There are however missing features, such as shell and pairing effects, but these feature are sub-dominant in the leptodermous expansion: they represent a refinement in the description of experimental binding energies which is of the order of a few MeV in total energy, compared to the leading order contributions which are of the order of hundred of MeV. The uncertainties are indeed still large at the leading order, as we will see in the following.

The contributions to the energy $E_\bulk$, $E_\fs$, $E_e$ and $E_{ng}$ will be detailed in the following sub-sections.

\subsection{The cluster bulk contribution}
\label{sec:bulk}

The cluster bulk contribution to the energy per particle is the leading order term in the leptodermous expansion (order $A_\cl$, the mass term). It is related to the homogeneous matter calculation, represented here by the MM energy density $\epsilon_\mm(n_n,n_p)=n_B e_\mm(n_n,n_p)$, given by
\begin{equation}
e_\bulk(I_\cl,n_\cl) = \frac 1 {n_B} \Big( \epsilon_\mm(n_n,n_p) - n_n m_n c^2 - n_p m_p c^2 \Big) \, ,
\label{eq:ebulk}
\end{equation}
where the neutron and proton masses $m_n$ and $m_p$ are fixed to their bare mass, $m_n c^2 = 939.565346$ MeV and $ m_p c^2 =  938.272013$~MeV, and $n_n$ and $n_p$ are the uniform neutron and proton densities in the cluster.

\subsection{The finite-size contribution}
\label{sec:FS}

The finite-size term in Eq.~\eqref{eq:enuc} incorporates the nuclear contributions to the cluster energy at all orders in the leptodermous expansion. In the present study, we limit ourself to the leading order terms: the Coulomb term is in $Z_\cl^2/A_\cl^{1/3}\approx A_\cl^{5/3}$ (dominant term at large $A$ which prevents super-heavy nuclei to exist), the surface is in $A_\cl^{2/3}$, and the curvature term is in $A_\cl^{1/3}$.
They are expressed as
\begin{eqnarray}
E_\fs(A_\cl,I_\cl,n_\cl) &=&  E_\coul(A_\cl,I_\cl) + E_\surf(A_\cl,I_\cl) \nonumber \\
&& + E_\curv(A_\cl,I_\cl),
\label{eq:efs}
\end{eqnarray}
where the Coulomb term for a spherical and uniform distribution of protons is given by the direct and exchange contributions,
\begin{eqnarray}
E_\coul &=& \mathcal{C}_\coul \left( E_{\coul, \dir} + E_{\coul, \exc} \right) \, ,
\label{eq:ecoul:tot}
\end{eqnarray}
with 
\begin{eqnarray}
E_{\coul, \dir} &=& \frac{3}{5}\frac{Z_\cl^2 e^2}{R_{p}} f_\coul(u) \, \\
&=& a_{c} \left(\frac{1 - I_\cl}{2}\right)^2 f_\coul(u) A^{5/3} \, \\
E_{\coul, \exc} &=& - \frac{3}{4}\left( \frac{3}{2\pi}\right)^{2/3}\frac{Z_\cl^{4/3}e^2}{R_{p}} h_\coul(u) \, \\
&=& - \frac{5a_{c}}{4}\!\left( \frac{3}{2\pi} \right)^{2/3} \!\! \left(\frac{1 - I_\cl}{2} \right)^{4/3} \!\!\!\!\! h_\coul(u) A \, ,
\end{eqnarray}
where $e^2\approx \hbar c/137$ and the functions $f_\coul$ and $h_\coul$ are defined as: $f_\coul(u)=1-(3/2)u^{1/3}+(1/2)u$ and $h_\coul(u)=1+u^{1/3}$, with $u$ the volume fraction of the cluster, defined as,
\begin{equation}
u = \frac{V_\cl}{V_\WS} = \frac{n_e}{n_{\cl,p}} = \frac{2 n_e}{(1-I_\cl) n_\cl} \, ,
\end{equation}
where $n_{\cl, p}=n_\cl (1-I_\cl)/2$ is defined as the proton density in the cluster. In the function $f_\coul$, the first term corresponds to the proton-proton repulsive interaction, the second term is the "lattice contribution" including the electron-proton and electron-electron interaction, under the hypothesis of a globally neutral Wigner-Seitz cell. Then the third term in $f_\coul$ stands for the finite-size correction which becomes important when the cluster volume is comparable with the Wigner-Seitz volume. This term is important near the crust-core transition and pushes the transition to nuclear matter towards higher densities. The first term is the only one remaining in the case of isolated nuclei, corresponding to the limit $u=0$. Since there is no proton-electron contributions to the exchange Coulomb energy, the first (second) term in $h_\coul$ corresponds to the proton-proton (electron-electron) contribution.

The coefficient $\mathcal{C}_\coul$ in Eq.~\eqref{eq:ecoul:tot} is a variational parameter which is fine tuned over the nuclear mass table. It describes -- in an effective way -- the effect the diffusive nuclear surface on the Coulomb energy, which is neglected in the sharp drop off density profile that we consider here. Since the diffusive surface is expected to be a small correction, the fit value is expected to remain close to 1, $\mathcal{C}_\coul\approx 1$.

Note that the direct Coulomb term scales like $A_\cl^{5/3}$ and therefore dominates the CLDM energy at large $A_\cl$. Since the Coulomb term is repulsive, this induces a limitation in the maximum $A_\cl$ for finite nuclei. However, for most of nuclei in the nuclear chart, the Coulomb interaction remains small compared to the nuclear one. For this reason, the bulk term in $A_\cl$ and the Coulomb direct contribution in $A_\cl^{5/3}$ are considered at the same order in the leptodermous expansion. It is also interesting to note that the exchange Coulomb term contributes to two order lower compared to the direct term in the $A_\cl^{1/3}$ leptodermous expansion. It is thus expected to effectively contribute at one order below the curvature contribution.

Neglecting the difference between the neutron and proton radii -- no skin approximation --, we have $R_p=R_\cl=r_{cl}(n_\cl)A_\cl^{1/3}$ with $r_{cl}^3(n_\cl)=3/(4 \pi n_{cl})$, and the Coulomb factor reads,
\begin{equation}
a_{c}(n_\cl) = \frac{3}{5} \left( \frac{4 \pi}{3} n_\cl  \right)^{1/3}e^2 \, .
\label{acfncl}
\end{equation}

Note that the Coulomb factor $a_c$ defined from Eq. (\ref{acfncl}) depends on the cluster density. The Coulomb parameter, $a_c$, is however often taken as a constant, see for instance Ref.~\cite{Carreau2019a}, either as a free parameter to be fitted or as function of the constant $n_{\sat}$. The different assumptions for $a_c$ give differences on the description of isolated nuclei. Note that if $a_c$ is taken to be constant (often taken to be of the order $\sim 0.7$ MeV), neither the Coulomb nor the surface energy contributes to nuclear pressure, the pressure derives from the bulk term only. In NS crust however, the Coulomb term contributes to the pressure thanks to its dependence in the volume fraction $u$. We show the difference of having $a_c$ constant or not in the sequence of the paper, FS1 refers to $a_c =  a_c(n_{\sat}) $, while in FS2 and others we have $a_c =a_c (n_{\cl})$.

\begin{table}[t]
\centering
\tabcolsep=0.27cm
\def\arraystretch{1.5}
\begin{tabular}{ccccc}
\hline\hline
$\sigma_\mathrm{surf,sat}$ & $\sigma_\mathrm{surf,sym}$ & $p_\surf$ & $\sigma_\mathrm{curv,sat}$ &  $\beta_\mathrm{curv}$ \\
MeV~fm$^{-2}$ & MeV~fm$^{-2}$ & & MeV~fm$^{-1}$ & \\
\hline
1.1 & 2.3 & 3.0 & 0.1 & 0.7 \\
\hline\hline
\end{tabular}
\caption{Standard FS parameters for the CLDM considered in this work. Note the associated value $b_\surf=29.9$.}
\label{table:stdParam}
\end{table}

The surface energy is proportional to the surface tension $\sigma_\surf(I_\cl)$ and scales as $A_\cl^{2/3}$. It reads
\begin{eqnarray}
   E_\surf(A_\cl,I_\cl,n_{\cl}) &=& 4 \pi R^2_\cl \sigma_\surf(I_\cl) \\ 
   &=&  4 \pi r^2_{\cl}  \sigma_\surf(I_\cl) A_\cl^{2/3} \, ,
   \label{eq:esurf}
\end{eqnarray}
with $\sigma_\surf(I_\cl)$ as expressed, as suggested in \cite{LattimerSwesty1991}, as
\begin{equation}
\sigma_\surf(I_\cl)=\sigma_{\surf, \sat} \frac{2^{p_\surf+1}+b_\surf}{Y_p^{-p_\surf}+b_\surf+(1-Y_p)^{-p_\surf}}, 
\label{eq:sigma}
\end{equation}
where $Y_p = Z_\cl/A_\cl = (1-I_\cl)/2$ is the cluster proton fraction and $\sigma_{\surf, \sat}$ is a parameter that determines the surface tension in symmetric nuclei. The parameter $p_\surf$ entering into the expression of the surface tension~\eqref{eq:sigma} plays an important role at large isospin asymmetries. It is usually fixed to be $p_\surf=3$ since the seminal contribution~\cite{LattimerSwesty1991}, but a small variation around $3$ plays an important role at large asymmetries, which occurs around the core-crust transition densities in NSs~\cite{Carreau2019b}.

For small asymmetries, 
\begin{equation}
\sigma_\surf(I_\cl)\approx \sigma_{\surf, \sat}-\sigma_{\surf,\sym} I_\cl^2
\label{eq:surfI}
\end{equation}
with
\begin{equation}
\sigma_{\surf,\sym} = \sigma_{\surf, \sat} \frac{2^{p_\surf} p_\surf(p_\surf+1)}{2^{p_\surf+1}+b_\surf}\, .
\end{equation}
One can thus relate the parameter $b_\surf$ to the surface symmetry energy $\sigma_{\surf, \sym}$.
We have
\begin{equation}
b_\surf = 2^{p_\surf} \left[ p_\surf (p_\surf+1) \frac{\sigma_{\surf, \sat}}{\sigma_{\surf,\sym}} - 2 \right] \, .
\end{equation}
In the following, we prefer to use the parameter $\sigma_{\surf, \sym}$ instead of $b_\surf$, since it directly reflects the isospin dependence of the surface tension for small isospin asymmetries, as shown in Eq.~\eqref{eq:surfI}. For this reason the domain of variation of $\sigma_{\surf, \sym}$ is better constrained than the one for the parameter $b_\surf$, which ease the determination of the prior for this parameter,
see section~\ref{sec:fit}.

Tab.~\ref{table:stdParam} suggests standard values for the FS parameters obtained by averaging over the usual parameters, see Ref.~\cite{Carreau2019b} for a sample of these parameters associated to various Skyrme interactions.

The curvature energy is controlled by the curvature tension $\sigma_{\curv}(I)$, and follows~\cite{Newton2012},
\begin{equation}
E_\curv(A_\cl,n_\cl,I_\cl) = 8 \pi r_{cl} \sigma_\curv(I_\cl) A_\cl^{1/3},
\end{equation}
with 
\begin{equation}
\sigma_\curv(I_\cl) = \alpha \sigma_{\curv,\sat} \frac{\sigma_\surf(I_\cl)}{\sigma_{\surf,\sat}}\left[\beta_\curv - \frac{1-I_\cl}{2}\right] \, .
\end{equation}
The parameter $\alpha$ is fixed to be $\alpha = 5.5$, since we allows the variation of the parameter $\sigma_{\curv,\sat}$ in the fit to the binding energy over the nuclear chart. The standard values for the curvature parameters $\sigma_{\curv,\sat}$ and $\beta_\curv$ are also given in Tab.~\ref{table:stdParam}.

\begin{table}[t]
\centering
\tabcolsep=0.2cm
\def\arraystretch{1.5}
\begin{tabular}{lccccc}
\hline\hline
Model          & Variables & FS1     & FS2     & FS3  &  FS4  \\
\hline
Bulk from MM      & ($I_\cl$, $n_{cl}$) & $\times$ & $\times$  & $\times$ & $\times$ \\
FS Surface        & ($n_{sat}$) & $\times$  & $-$  & $-$ & $-$ \\
FS Coulomb (Dir.) & ($n_{sat}$) & $\times$  & $-$ & $-$ & $-$ \\
FS Surface        & ($n_{cl}$)  & $-$       & $\times$  & $\times$ & $\times$ \\
FS Coulomb (Dir.) & ($n_{cl}$)  & $-$       & $\times$  & $\times$ & $\times$ \\
FS Curvature      & ($n_{cl}$)  & $-$   & $-$  & $\times$  & $\times$ \\
FS Coulomb (Ex.)  & ($n_{cl}$)  & $-$ & $-$  &  $-$ & $\times$ \\
\hline
Number of param. & & 3  & 3    & 5  & 5  \\
\hline\hline
\end{tabular}
\caption{Definition of nuclear macroscopic models used in this work. In the left column we have the different terms on the CLDM implemented in this work. On the first line we have the label of the model. The table show which term and how many parameters are necessary to fix each model.}
\label{table:FSmodels}
\end{table}

In the present paper, we explore the role of various approximations in the FS terms on the NS crust properties. We order these approximations by their expected impact and selected four of them, that we call FS1 to FS4, see Tab.~\ref{table:FSmodels}:
\begin{itemize}
\item FS1 is the simplest approximation for the FS term that we consider, where only the surface and direct Coulomb contributions are included and for which the Coulomb and surface parameters, which are related to $r_\cl$, are taken constant and fixed by setting $n_\cl=n_\sat^\emp$.
\item FS2 is an improved version of FS1, where the Coulomb and surface parameters are varied and fixed from the actual value of the cluster density $n_\cl$. 
\item In FS3, we additionally incorporate the effect of the curvature contribution.
\item Finally in FS4, we add the exchange Coulomb contribution to the Coulomb energy term.
\end{itemize}

\begin{figure*}[t]
\centering
\includegraphics[scale=0.34]{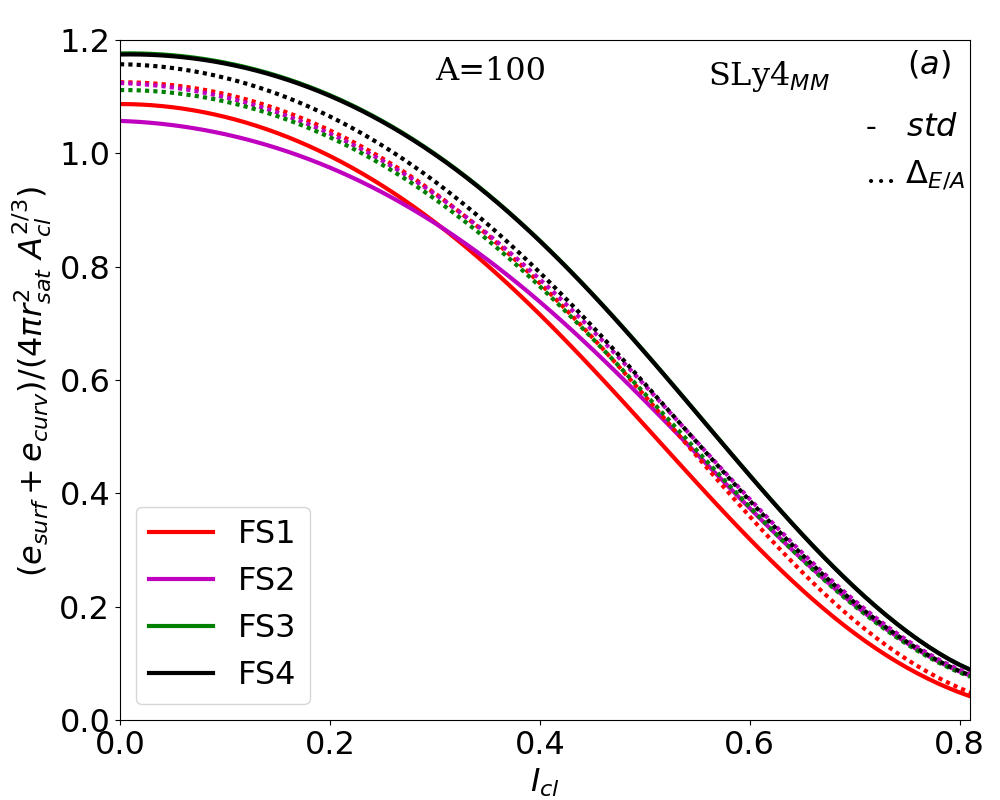}
\includegraphics[scale=0.34]{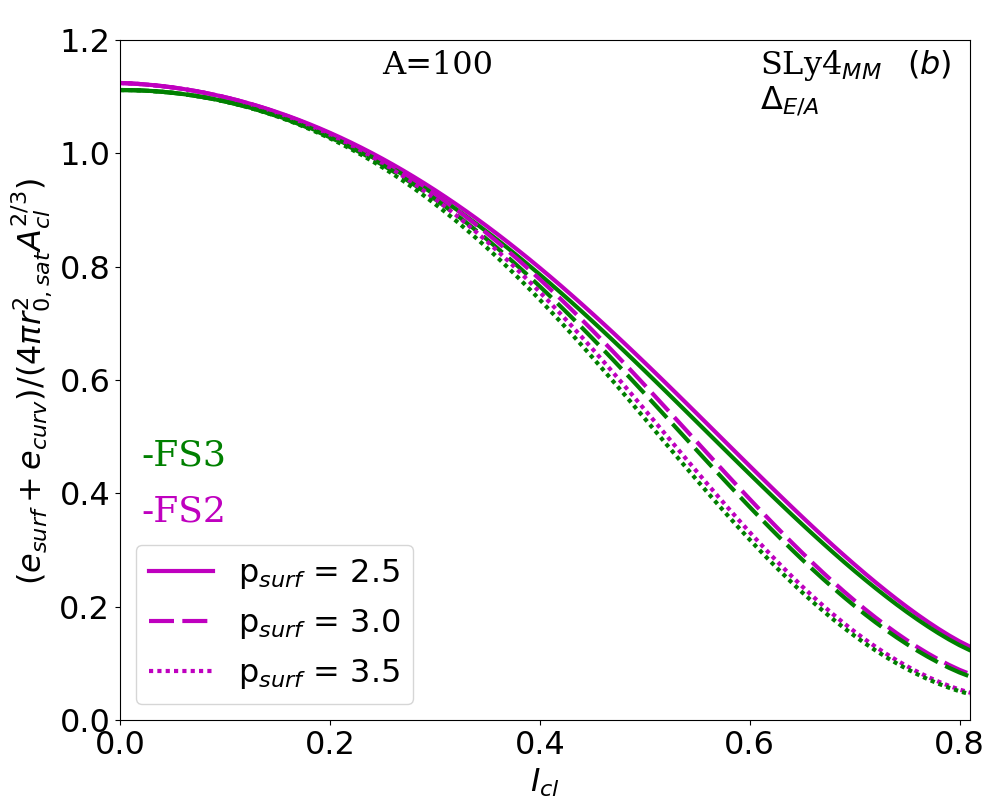}
\caption{Sum of surface and curvature energies normalized by $4\pi r_{0,\sat}^2 A_\cl^{2/3}$ function of the cluster asymmetry $I_\cl$ in isolated nuclei. The left panel shows a comparison between the four finite-size models, in the case they are fixed by the standard values given in Table~\ref{table:stdParam} (solid lines) and in the case they are fit to the nuclear masses (dotted lines). The right panel shows the influence of the surface parameter $p_\surf$ at large $I_\cl$, after the fit to the nuclear masses.}
\label{fig:EsurfFS}
\end{figure*}

In the future, we plan to incorporate more contributions and go beyond FS4. The approximations captured into the FS1-4 models represent however a systematical development where the refinements are expected to play a smaller and smaller role. This will be confirmed in Sec.~\ref{sec:NS}. The FS1-4 models allows also to understand the differences between the models proposed for the crust. In their seminal paper in 1971, Bethe, Baym, and Pethick~\cite{bbp1971} have introduced the first version of the CLDM with Coulomb and surface terms only, similar to our FS2 with the addition of neutron skin effect and without optimization to the nuclear chart. Another well-known model for the crust was proposed by Douchin and Haensel in 2001~\cite{DouchinHaensel2001}, still considered a CLDM model with the additional contribution of the curvature term, as in our model named FS3. The surface and curvature terms were however determined from many-body methods~\cite{DouchinHaensel2000}. They have also considered different geometries in the pasta phase and incorporated the effect of skins in the CLDM, which go beyond the present approximations. In 2008, Steiner considered a CLDM~\cite{Steiner2008} with surface and Coulomb terms comparable to our FS2 approximation, but introduced in addition an effective way to describe skins that we have previously discussed. Newton \etal suggested in 2012~\cite{Newton2012} a CLDM with a full Coulomb term (direct and exchange) as well as surface and curvature contributions, as our FS4 approximation. They additionally studied different geometries in the core-crust region (pasta phases). In 2016, Tews adopted the same CLDM as the one suggested by Steiner~\cite{Steiner2008} and he additionally considered a bulk contribution determined by chiral EFT calculations~\cite{Tews2017}. The FS terms were also adjusted to reproduce the binding energy over the nuclear chart. These two features make this study comparable to our model for the crust. In 2017, Vi\~nas \etal~\cite{Vinas2017} extended the CLDM by considering the Thomas-Fermi approximation and introduced surface, Coulomb and curvature terms, all FS terms being function of the proton radius. This treatment is similar to our FS4 approximation, but with additional contribution due to skin and pasta phases. In 2019, Carreau \etal use a CLDM with surface and Coulomb terms in \cite{Carreau2019a}, which can be compared to our FS1 approximation, and then included in 2020 curvature and shell effects~\cite{Carreau2020}, which can be compared to our FS3, but fixing $a_c$ constant and including shell effects.

Fantina \etal \cite{Fantina2013} constructed a unified EOS within Brussels-Montreal Skyrme (BSk) functionals, using experimental nuclear masses at the outer-crust, with the Brussels-Montreal mass model when there is no data, and a  extended Thomas-Fermi plus Strutinsky integral method at the inner crust \cite{Pearson2012}. This model is an interesting alternative to describe neutron star crust, where it is possible to model shell effects. Recently this approach inspired a new model combining the advantage of CLDM with the modelling of shell effects ~\cite{Carreau2020}.

In conclusion, there is not a strict equivalence between the FS1-4 approximations we suggest and the various models for the crust which have been investigated, but we believe our series of approximation is incremental and well suited to the understanding of the role of various terms, and associated uncertainties, on the properties of the crust of NS. The comparison to other crust models suggests also that our FS approximation series could be extended in order to incorporate neutron skins, shell effects, and different geometries in the core-crust region.

\begin{figure*}[t]
\centering
\includegraphics[scale=0.34]{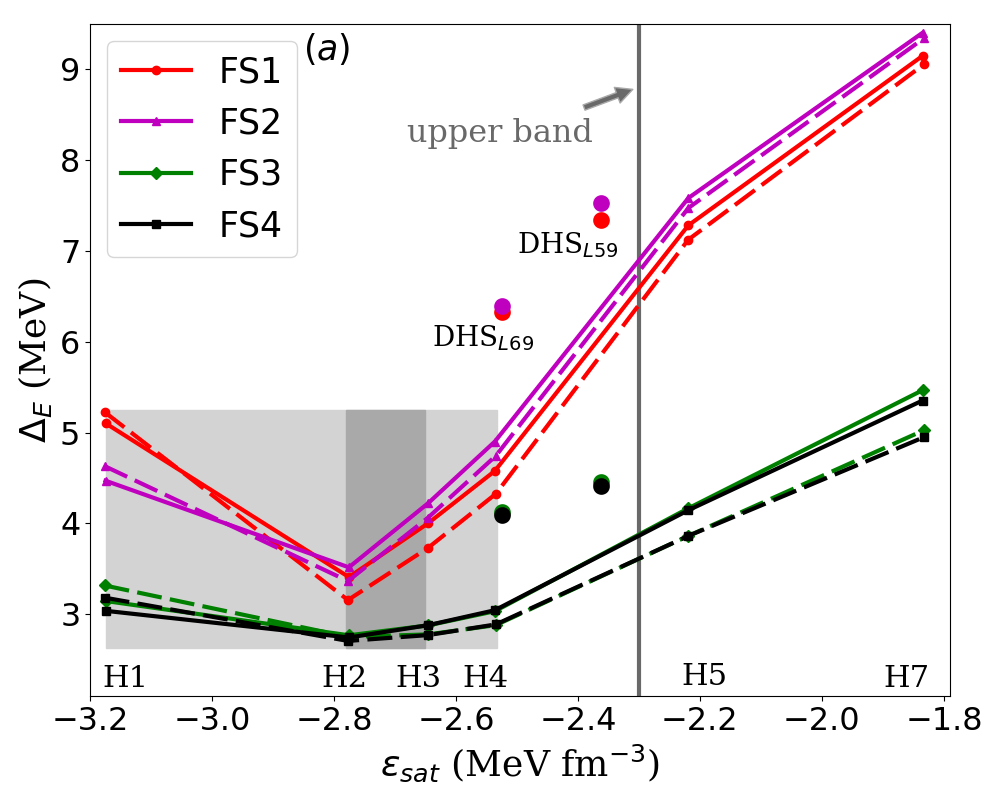}
\includegraphics[scale=0.34]{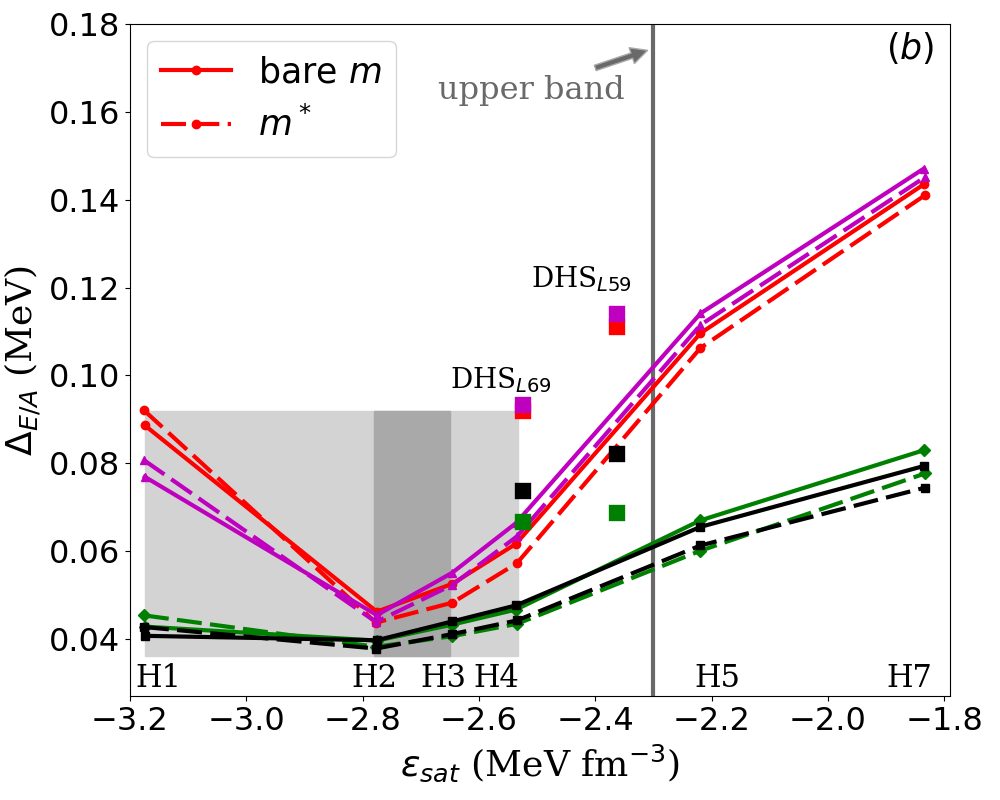}
\caption{Loss function $\Delta_E$ fitted to the total energy (left) and $\Delta_{E/A}$ fitted to the energy per particle (right) w.r.t $\epsilon_\sat$  for FS1-FS4 for H1-H5 and H7 (DHS$_{L59}$ and DHS$_{L69}$) in lines (symbols). Solid lines stand for the bare mass while dashed lines for the effective mass.}
\label{fig:loss}
\end{figure*}

\subsection{Fit to experimental nuclear masses}
\label{sec:fit}

In this section we discuss the impact of the fit on isolated nuclei. We analyse two definitions for the loss function, $\Delta_{E}$ or $\Delta_{E/A}$, which are defined as:
\begin{eqnarray}
\Delta_{X} &=& \left[ \frac{1}{N} \sum_{i=1}^{N} ( X_\ex^i - X_\nuc^i )^2 \right]^{1/2}\, , 
\label{eq:loss}
\end{eqnarray}
where $X=E$ or $X=E/A$: $X_\ex^i$ is the experimental values and $X_\nuc^i$ is the energy given by the CLDM model. In the present fit, we have considered $N=3375$ nuclei taken from the 2016 Atomic Mass Evaluation (AME) \cite{AME2016}, with nuclei in the range: $A = 12 - 295$ and $Z = 6 - 118$. 

It is interesting to analyze the overall impact of the fine tuning of the mass models on the surface and curvature contributions to the nuclear energy. To do so, we represent in Fig.~\ref{fig:EsurfFS}(a) the following quantity $(e_\surf + e_\curv)/(4\pi r_{\sat}^2 A_\cl^{2/3})$ as function of the nuclear asymmetry $I_\cl$ for a typical value $A_\cl=100$. The homogeneous contribution is also fixed to be given by SLy4$_\mm$. The different colors shows the result of the different FS terms, as indicated in the legend, the solid lines correspond to the standard values, given in Tab.~\ref{table:stdParam}, while the dashed lines results from the minimization of the loss function $\Delta_{E/A}$. The overall trend is similar for all the considered cases.  The contribution of the curvature term for FS1 and FS2 is null. The difference between FS1 and FS2 reflects the role played by the parameter $r_{\cl}^2$ in FS2 while it is fixed to be $r_{\sat}^2$ in FS1: since the cluster density $n_\cl$ decreases as $I_\cl$ increases, $r_{\cl}^2$ becomes larger than $r_{\sat}^2$ at large $I_\cl$, as shown in  Fig.~\ref{fig:EsurfFS}(a). The consistent treatment of the cluster radius in the FS terms, included in FS2 and further approximation, tends to increase the surface term at large asymmetry $I_\cl$ compared to FS1, where the values radius parameter is constant.
 
Employing the standard parameter set (solid lines), FS3 and FS4 predict larger values than FS1 and FS2, since the contribution of the curvature $e_\curv$ adds up to the surface one $e_\surf$. The exchange Coulomb term has no effect on the surface tension, therefore FS3 and FS4 shows identical curves in the figure. The dispersion between the dotted lines is smaller than between the solid lines as a result of the minimization which is performed for each FS model. Finally, in the fit the curvature contribution is absorbed by a reduction of the surface one, such that the sum remains identical, as seen from the comparison of FS3 to FS4 (dashed lines). The fit with the Coulomb exchange in FS4, absent in FS3, tends to slightly increase the surface and curvature terms around $I_\cl\approx 0$.

Fig. \ref{fig:EsurfFS}(b) shows the influence of the surface parameter $p_\surf$ on the same quantity as in Fig. \ref{fig:EsurfFS}(a), where we show only FS2 and FS3 for clarity. Around $I_\cl\approx 0$ the impact  of varying $p_\surf$ is null, while it plays an major role for $I_{\cl} > 0.3$, as it was already remarked in Ref.~\cite{Carreau2019a}. As a consequence $p_\surf$ cannot be determined from the confrontation to the experimental nuclear chart~\cite{Carreau2019a}. It however plays an important role in the densest layers of the NS crust at the vicinity of the core, where matter is the most neutron rich. The value of the parameter $p_\surf$ is thus an important source of uncertainties which cannot be controlled by the nuclear experimental data.
This will be illustrated in Sec.~\ref{sec:psurf}.

We now vary both the bulk and the FS terms, where the bulk terms we consider are the ones which reproduce the MBPT predictions based on chiral interactions, see Sec.~\ref{sec:homogeneous}. The impact of varying the bulk contribution on the minimization based on either the two loss functions $\Delta_E$ and $\Delta_{E/A}$ is shown in Fig.~\ref{fig:loss}. The horizontal axis is chosen to be $\epsilon_\sat=n_\sat E_\sat$ and the different colors represent the best fit obtained for each FS approximation (FS1 to FS4). The solid (dashed) lines are obtained with the bare (effective) nucleon mass. There is a small improvement using the effective mass instead of the bare mass, but it however remains small compared with the impact of changing the Hamiltonian.

\begin{figure*}[t]
\centering
\includegraphics[scale=0.35]{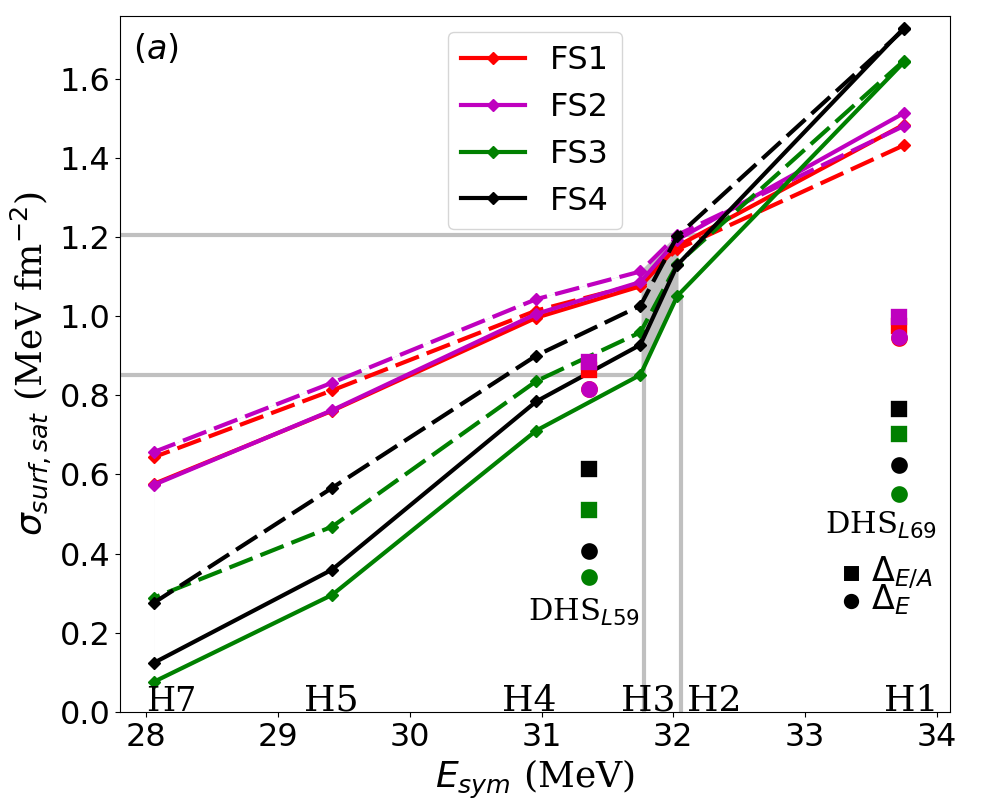}
\includegraphics[scale=0.35]{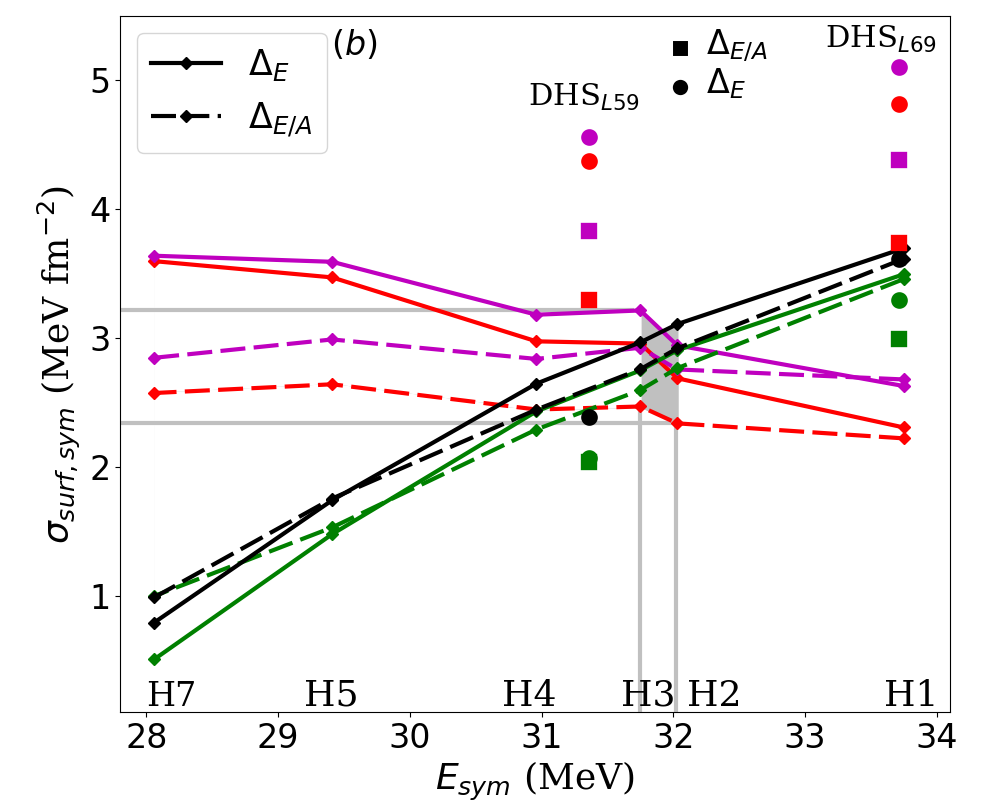}
\caption{Top: The isoscalar surface tension parameter $\sigma_{\surf, \sat}$ is represented against $E_\sym$ for the six Hamiltonians H1-H5 and H7, and for the FS models FS1 to FS4. The two Hamiltonians DHS are included with dots (squares) for the minimization to the total energy (energy per particle). Bottom: Same for $\sigma_\sym$. Silver band shows the values for the two Hamiltonians which best reproduce nuclear masses, H2 and H3. }
\label{fig:surface}
\end{figure*}

For instance, the Hamiltonians for which $\epsilon_\sat>-2.30$~MeV~fm$^{-3}$ are less good in reproducing the experimental nuclear masses than the others. The reason is that the reproduction of the experimental binding energies requires a delicate balance between the attractive bulk term $\epsilon_\sat$ and the repulsive FS term $\epsilon_\fs$. As $\epsilon_\sat$ increases, the FS term decreases such that the sum remains constant. The compensation can happen until the FS term becomes almost zero. Above this limit, if $\epsilon_\sat$ continues to increase the energy density becomes too large to be able to reproduce the experimental masses, and the quality of the fit gets worst and worst, as illustrated in Fig.~\ref{fig:loss}. 

The groups FS3-4 -- where the curvature term has been incorporated -- reduce the loss function compared to the groups FS1-2.
The best models are H1, H2, H3, and H4, among which H2 and H3 are even better. In the following analysis, the predictions based on H1-H4 will be marked with a light grey band, while the ones based on H2-H3 will be identified with a darker grey band, as illustrated in Fig.~\ref{fig:loss}.

We also analyse in Fig.~\ref{fig:loss} the CLDM based on the Hamiltonians DHS. Note that it is still $\epsilon_\sat$ which drives the goodness of the model: the closer it is to -2.6/-2.8~MeV~fm$^{-3}$ the better is the agreement with experiments. There is however a reduction of the goodness of these models as the value of $L_\sym$ departs from the one of the other models by 10 to 15~MeV, see Table~\ref{tab:empirical:Hamiltonians}. Since the DHS Hamiltonians do not show a good reproduction of experimental masses, we neglect them on the NS crust study. In the next section we show the Hamiltonians which better reproduced nuclear masses, H1-H4, in gray band together with H5 and H7 for comparison. 

Finally, we explore the correlation between the surface energy parameters $\sigma_{\surf, \sat}$ and $\sigma_{\surf, \sym}$ and the symmetry energy $E_\sym$ for the Hamiltonians H1-H7, the FS models FS1-FS4, the prescription of the nucleon mass, and the loss function used for the minimization, as shown in Fig.~\ref{fig:surface}. Having or not the curvature contribution in the CLDM is the main source of difference between these correlations. With the contribution of the curvature term the parameters $\sigma_{\surf, \sat}$ and $\sigma_{\surf, \sym}$ are almost linearly correlated with $E_\sym$. The correlation between $\sigma_{\surf, \sym}$ and $E_\sym$ was already discussed in Ref.~\cite{Steiner2008}, but we note here that this correlation is model dependent: without the curvature term it become an anti-correlation, which can even be almost flat if the minimization of the mass table is based on the loss function $\Delta_{E/A}$.

It is further interesting to note that even if these correlations are model dependent, they are crossing for the value of the symmetry energy around 32~MeV, comparable to the one of the Hamiltonians H2 and H3. Therefore, selecting H2-H3 Hamiltonians reduces strongly the model dependence of the parameters $\sigma_{\surf, \sat}$ and $\sigma_{\surf, \sym}$. In the future, it will be interesting to check if this property remains while adding more terms to the FS approximation.

The conclusion of this confrontation of the CLDM to the experimental nuclear masses is that the Hamiltonians H1-H4 which satisfy the condition $\epsilon_\sat<-2.30$~MeV~fm$^{-3}$ reproduce well the experimental masses over the mass table. A better reproduction over the mass table is obtained for the Hamiltonians for which $\epsilon_\sat=-2.70(20)$~MeV~fm$^{-3}$. We obtained the best results for the confrontation of the CLDM to the experimental masses for the Hamiltonians H2 and H3, which will represent in the following the best models for the NS crust properties.

\subsection{The electron and neutron gas contributions}
\label{sec:gas}

At variance with isolated nuclei, the energetics of the NS crust incorporates the contributions from the electron and neutron gas, that we briefly present here for completeness.

Similarly to the contribution to the bulk properties for the nuclear clusters, the energy of the neutron gas is given by the MM as,
\begin{equation}
e_{ng}(n_{ng}) = \frac{1}{n_{ng}} \epsilon_\mm(n_{ng} = n_n,n_p=0) \, .
\end{equation}

Note that by considering the same model providing the core properties and the bulk and gas contribution in the crust, our CLDM provides a unified description of the entire NS~\cite{DouchinHaensel2001,Fantina2013,Fortin2016}.

Since the electron interaction between electrons and between electrons and protons have already been absorbed in the Coulomb term, the remaining electron gas contribution is purely kinetic and reads for a relativistic gas,
\begin{eqnarray}
\epsilon_{e} &=& C_e \left[ x_e (1+2x_e^2)\sqrt{1+x_e^2} - \mathrm{asinh}(x_e) \right] \,.
\end{eqnarray}
where $x_e = \hbar c \; k_{F_e} / m_ec^2$, the electron Fermi momentum $k_{Fe}=(3\pi^2 n_e)^{\frac{1}{3}}$, and the overall constant $C_e=(m_e c^2)^4/(8 \pi^2 (\hbar c)^3)$. The electron chemical potential is expressed as,
\begin{equation}
\mu_{e} = m_e c^2 \sqrt{ 1+x_e^2 } \, ,
\end{equation}
and the pressure,
\begin{eqnarray}
P_{e}&=& - \epsilon_{e} + n_e \mu_e \, .
\end{eqnarray}

\section{NS crust properties}
\label{sec:NS}

In this section, we derive the equilibrium configurations in the crust of NS described by the CLDM~\eqref{eq:ecl}. We first derive the equilibrium equations and then present and discuss our results for the NS equation of state.

\subsection{Equilibrium equations in the crust}

The CLDM cluster energy~\eqref{eq:ecl} is minimized under the constraint of the baryon density $n_B$ defined as,
\begin{eqnarray}
n_B &=& \frac{A_\cl+N_{g}}{V_\WS} = n_\cl u + n_{ng} (1-u) \, , \\
 &=& n_{ng} + \frac{2 n_e}{1-I_\cl}\left( 1 - \frac{n_{ng}}{n_\cl}\right) \, .
\label{eq:nb21}
\end{eqnarray}
In Eq.~\eqref{eq:nb21}, the density $n_B$ is expressed in terms of four of the five independent variables: $A_\cl$, $I_\cl$, $n_\cl$, $n_e$ and $n_{ng}$. We use the Lagrange multipliers technique, as suggested in Ref.~\cite{Gulminelli2015}, to minimize the canonical energy density $\epsilon_\mathrm{can}$ in the Wigner-Seitz cell, 
\begin{equation}
\epsilon_\mathrm{can} = \epsilon_\tot - \mu_B n_B \, .
\label{eq:ecan}
\end{equation}
The total energy reads,
\begin{equation} 
\begin{split}
    \epsilon_\tot(A_\cl,I_\cl,n_{cl},n_e,n_g) = u  \epsilon_{\cl}+
    \left (1 - u  \right ) { \epsilon}_g     \\
   +n_e(m_p c^2 - m_n c^2)+n_B m_n c^2  +\rho_e\, ,
\end{split}
\label{eq:eCrust}
\end{equation}
where $u = V_\cl /V_\WS$ is the volume occupied by a nucleus in a Wigner-Seitz cell, $\rho_e$ is the electron energy density (with rest mass), $\epsilon_g = \epsilon_\mm (n_{ng}=n_g,n_p=0)$ is the neutron gas energy density and $\epsilon_\cl =  \epsilon_\mm (n_{n,\cl},n_{p,\cl}) + \epsilon_\fs(A_\cl,I_\cl,n_e,n_\cl)$ is the cluster energy density, with the finite-size contributions given by Eq. \eqref{eq:efs} and discussed in the previous section. 

In fact $\epsilon_\mathrm{can}$ coincides with the canonical potential, $\epsilon_\mathrm{can}=-P_\tot$. So minimizing $\epsilon_\mathrm{can}$ is equivalent to maximizing the total pressure $P_\tot$. Moreover, minimizing the total energy $E_\tot\equiv \epsilon_\tot V_\WS$ at fixed baryon density $n_B$ is equivalent to minimizing the total Gibbs energy $G_\tot$ at fixed total pressure, as discussed in Ref.~\cite{Pearson2012}.

\begin{figure*}[t]
\centering
\includegraphics[scale=0.485]{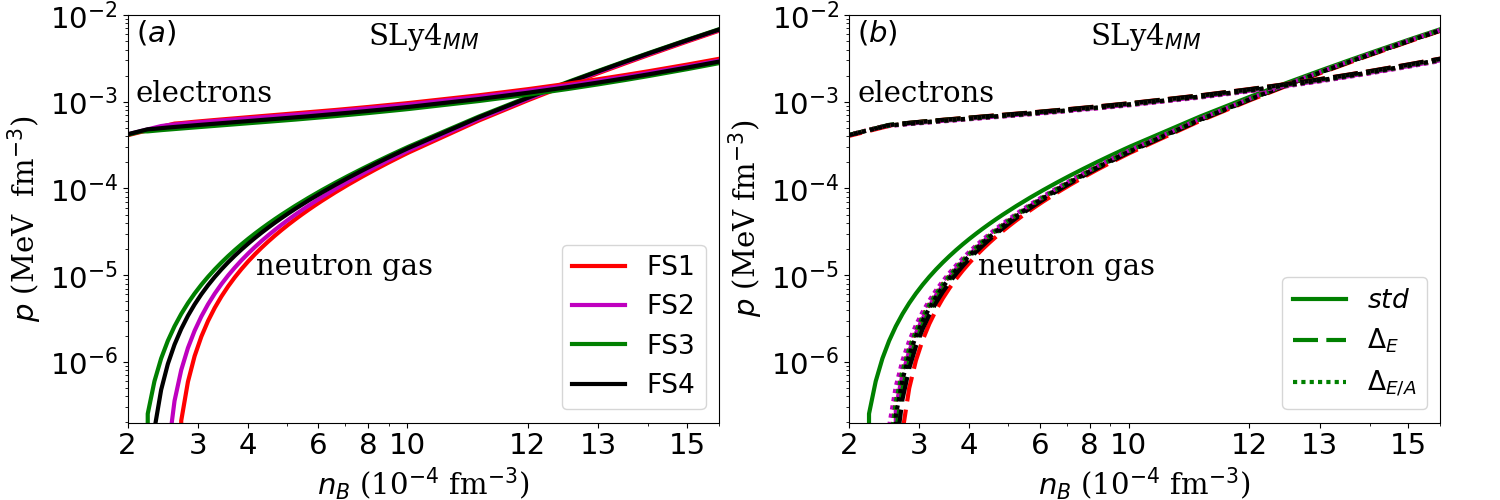}
\caption{Neutron gas and electron contribution to the pressure on the NS crust. Left panel shows the difference originated by the four finite-size models (with standard parameters, see Table \ref{table:stdParam}). Right panel shows the influence of the different optimization procedures (see Section \ref{sec:fit}).}
\label{fig:pressFS}
\end{figure*}

We define the following thermodynamical quantities ($q=n$, $p$),
\begin{eqnarray}
P_\cl &\equiv& n_\cl^2 \frac{\partial E_\cl/A_\cl}{\partial n_\cl}\bat_{A_\cl,I_\cl} \label{eq:Pcl} \, , \\
P_g &\equiv& -\epsilon_g + n_g \mu_g \label{eq:Pg}\, ,\\
\mu_{\cl, q} &\equiv& \mu_{\nuc, q} + \frac{P_g}{n_B} \label{eq:mucl}\, , \\
\mu_e &\equiv& \frac{\partial E_e}{\partial N_e}\bat_{N_\cl, Z_\cl} +\frac{2 n_e}{(1-I_\cl)A_\cl} \frac{\partial E_\coul}{\partial n_e} \label{eq:mue}\, , 
\end{eqnarray}
with
\begin{equation}
\mu_{\nuc, n} \equiv \frac{\partial E_\nuc}{\partial N_\cl}\bat_{Z_\cl, N_e} \, , \quad
\mu_{\nuc, p} \equiv \frac{\partial E_\nuc}{\partial Z_\cl}\bat_{N_\cl, N_e} \, .
\end{equation}
From Eq.~\eqref{eq:Pcl}, we deduce,
\begin{equation}
P_\cl = P_\bulk + P_\coul + P_\surf + P_\curv \, ,
\end{equation}
with obvious definitions for these partial contributions.

We impose the stationary of the canonical potential \eqref{eq:ecan} with respect to the five independent variables, considering $\mu_B$ as a constant parameter, and obtain the following equilibrium relations~\cite{Carreau2020},
\begin{eqnarray}
2E_\coul &=& E_\surf + 2 E_\curv \, , \label{eq:virial}\\
P_\cl &=& P_g \, , \label{eq:meca}\\
\mu_{\cl, n} &=& \mu_g \, , \label{eq:chemical}\\
\mu_{\cl, n} &=& \mu_{\cl, p} + \mu_e + \Delta mc^2 \, , \label{eq:beta}\\
\mu_B &=& \mu_g \label{eq:pot} \, ,
\end{eqnarray}
where $\Delta m=m_p-m_n$.
These equilibrium relations have a physical understanding: Eq.~\eqref{eq:virial} is an extension of the virial theorem~\cite{Baym1971} including the curvature energy, Eq.~\eqref{eq:meca} reflects the mechanical equilibrium, Eq.~\eqref{eq:chemical} describes the chemical equilibrium between the cluster and the gas in the $r$-representation, and finally Eq.~\eqref{eq:beta} assures the $\beta$-equilibrium. The last Eq.~\eqref{eq:pot} describes the baryon chemical potential which fixes the baryon density.

Equations \eqref{eq:virial}-\eqref{eq:pot} are solved by using the robust Newton-Rhapson method. Note that since the surface energy~\eqref{eq:esurf} is independent of the gas density $n_{ng}$, Eq.~\eqref{eq:pot} is subsumed to $\mu_B=\mu_g$.

In the following, we analyse the role played by the different terms in the CLDM and discuss their role in the uncertainties on the various observables in the NS crust.

\subsection{Impact of the FS terms and of the loss function}

We first discuss the role of the FS terms and of the loss function, see Eq.~\eqref{eq:loss}, used for the confrontation to experimental nuclear masses.

The neutron gas $P_g$ and the electron $P_e$ pressures are shown in Fig.~\ref{fig:pressFS} as function of the baryon density $n_B$. We remind that the equilibrium condition in the WS cell imposes that $P_g=P_\cl$, and the total pressure is $P_\tot=P_g+P_e$. Fig.~\ref{fig:pressFS}(a) shows the impact of the FS terms by fixing the FS parameters to their standard values (without optimization), see Table \ref{table:stdParam}, and the bulk term is derived from SLy4$_\mm$. The neutron gas pressure is largely impacted by the FS approximation at low density. The differences between the curves reflects the different density at which the outer-inner crust phase transition takes place. The main differences are between the groups FS1-FS2 and FS3-FS4, reflecting the important role of the curvature term. As the density increases, the effect of FS terms on the cluster pressure is weakened. Note that at low densities the electron pressure dominates over the neutron gas pressure. As a consequence, the total pressure will be weakly impacted by the FS terms, but the equilibrium configurations at the bottom of the inner crust could potentially be.

Fig.~\ref{fig:pressFS}(b) is similar to Fig.~\ref{fig:pressFS}(a), but we show the electron and gas pressure after the optimization of the FS parameters, i.e. including the impact of the experimental nuclear masses. The dashed (dotted) lines stand for the minimization employing the loss function $\Delta_E$ ($\Delta_{E/A}$). We observe that the optimization of the FS parameters on the experimental nuclear masses considerably reduces the dispersion between the various FS terms, as shown in Fig.~\ref{fig:pressFS}(a). There are still some differences, but they are smaller once the experimental nuclear masses are considered. For instance, the transition density between the outer and the inner crust is better determined. The impact of the loss function, while not negligible, is also small compared to the uncertainties from the FS terms.

\begin{figure*}
\centering
\includegraphics[scale=0.45]{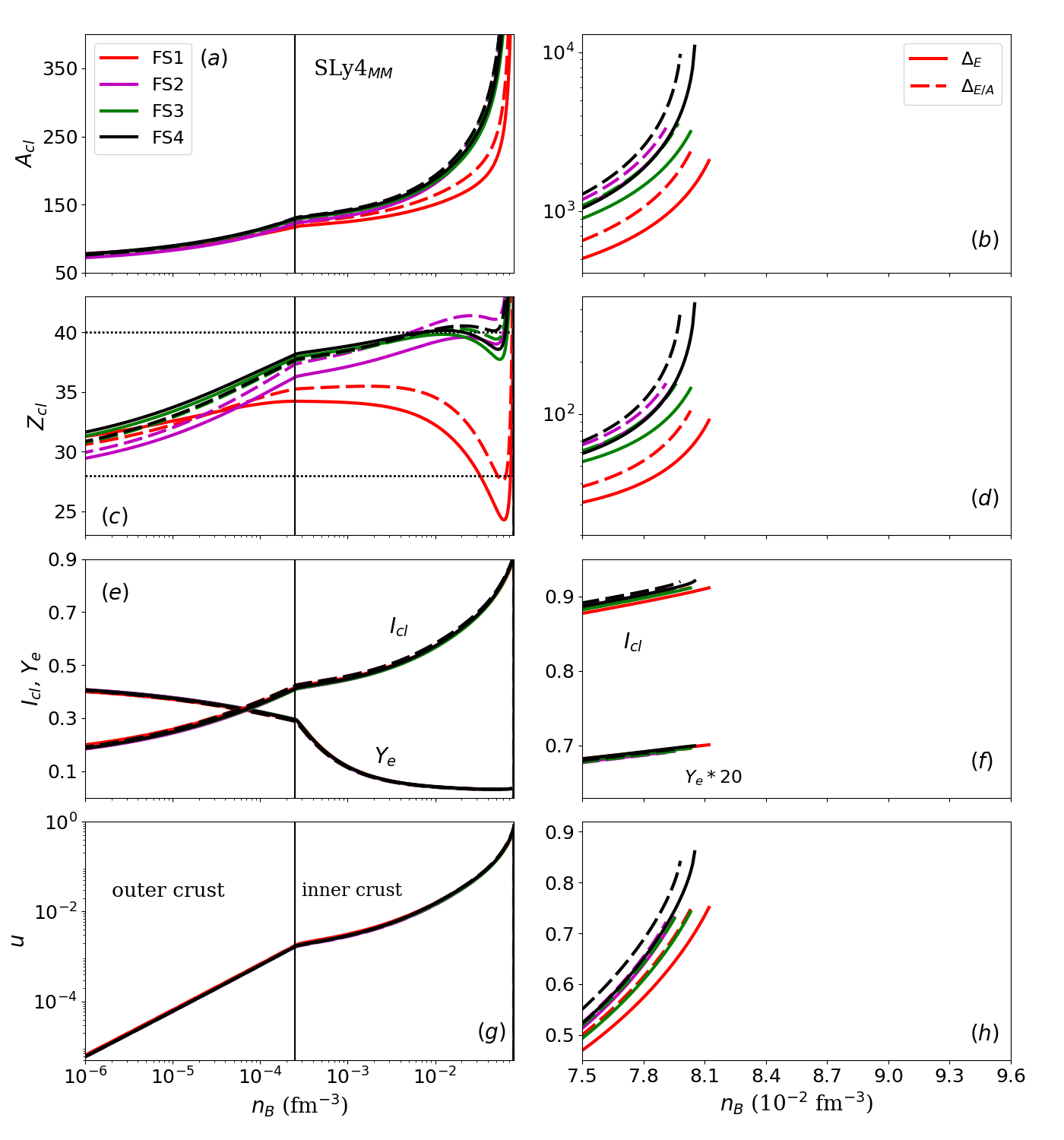}
\caption{Crust composition (top), asymmetry (center) and volume fraction occupied by the cluster (bottom) for the different ingredients in the CLDM within SLy4$_\mm$ interaction. FS1 (red), FS2 (magenta), FS3 (green) and FS4 (black) as explained in Tab.~\ref{table:FSmodels}. Dotted lines on panel (c) represents the magic numbers $Z_\cl=28$ and $Z_\cl=40$. Continuous line represent the parameters fitted to the total energy, while dashed lines represent the fitting to the energy per particle. Left: outer and inner crust. Right: zoom at crust-core transition.}
\label{fig:FStermsSLy4}
\end{figure*}

We now analyze more globally the properties of the crust -- from outer to inner -- in terms of composition ($A_\cl$ and $Z_\cl$), isospin asymmetries ($I_\cl$ and $Y_e$) and volume fraction $u$, see Fig.~\ref{fig:FStermsSLy4}. Here the bulk term is still SLy4$_\mm$. Note that the right panels (b, d, f, h) shows the same quantities as the left panels (a, c, e, g), but zooming in the crust-core transition region. The end of the curves indicates the boundary of the inner crust where the phase-transition to uniform matter occurs. The crust-core transition is defined as the density at which the energy density in the crust given by Eq.~\eqref{eq:eCrust} matches with $e_\mm$ the energy density in uniform matter given from the MM at the same density.

\begin{figure*}
\centering
\includegraphics[scale=0.45]{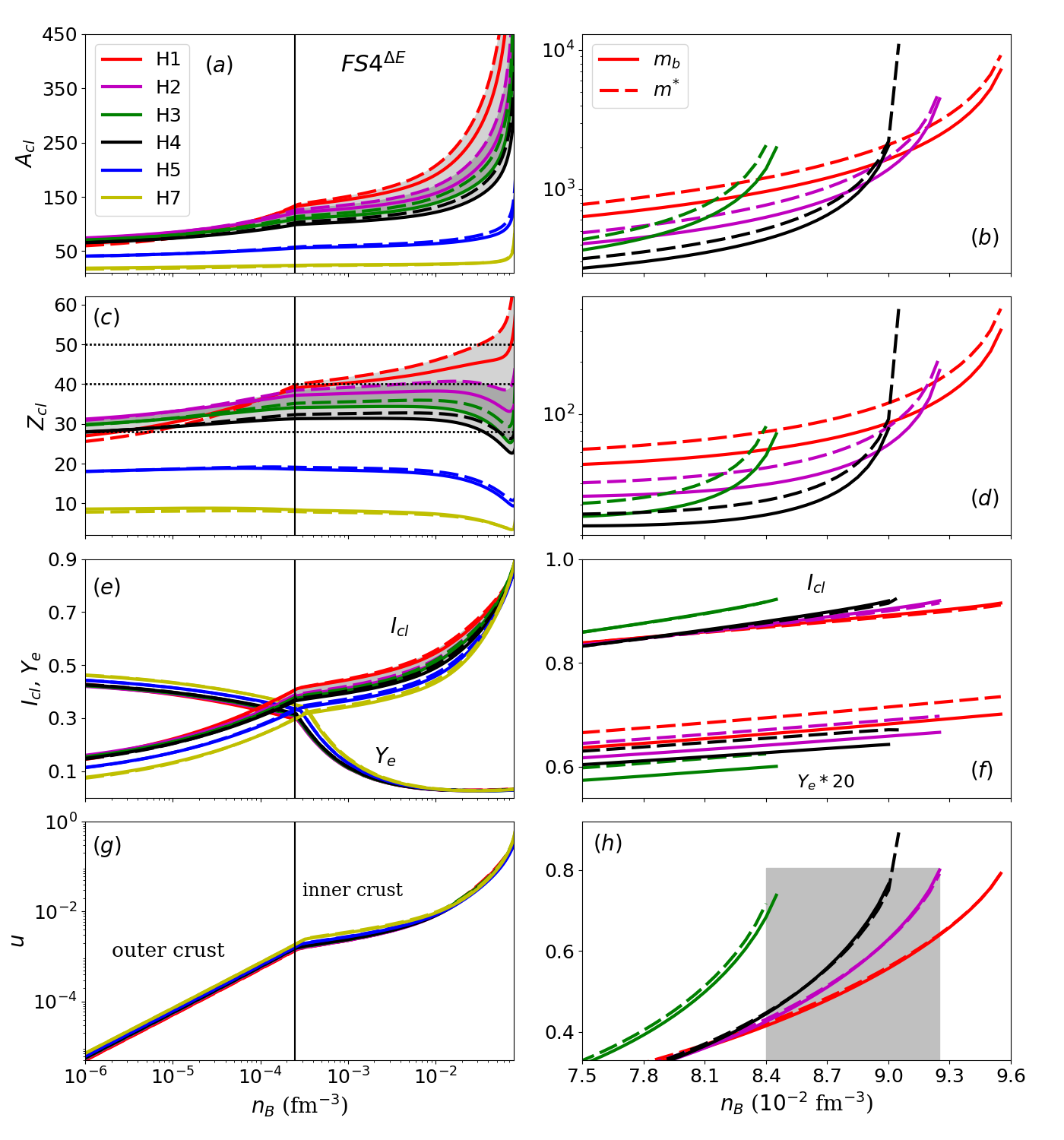}
\caption{Crust composition (top), asymmetry (center) and volume fraction occupied by the cluster (bottom) for the different ingredients in the CLDM. In the left pannels we show the 6 H  from outer to inner crust, on the right we show the 4 H allowed by the finite nuclei analysis with a zoom close the crust-core transition. Dotted lines on panel (c) represents the magic numbers $Z_\cl=28$, $Z_\cl=40$ and $Z=50$. Dashed lines includes the nucleon effective mass. All figures use FS4$^{\Delta_E}$ model.}
\label{fig:crust7H-FS4baremass}
\end{figure*}

As expected, there is a hierarchy in the contributions of the FS terms -- from FS1 to FS4 -- since the largest difference is observed between FS1 and FS2, then between FS2 to FS3, and finally from FS3 to FS4. The different observable are however not identically sensitive to the FS terms. For instance $A_\cl$, $Z_\cl$, and $I_\cl$ are impacted, while $Y_e$, $u$ and $n_\cc$ (the crust-core transition density) are almost insensitive to the FS terms. For reference, the magic number $Z_\cl$=28 and the quasi-magic number $Z_\cl$=40 are shown in Fig.~\ref{fig:FStermsSLy4}(c). The number of protons $Z_\cl$ in the inner crust is more stable employing FS4, which includes curvature and exchange Coulomb terms, as well as a proper treatment of the cluster density $n_\cl$ in the equilibrium equations, compared to the lower order FS terms. We will see in the following that this stability is also due to the value taken of $p_\surf$. The impact of the loss function is also minimal for FS4, compared to the other FS terms. The composition of the neutron star crust is important for the determination of transport properties  \cite{Strohmayer1991,Schmitt2018}, which has a direct impact on shear frequencies of neutron stars \cite{Tews2017}.

\begin{figure*}
\centering
\includegraphics[scale=0.45]{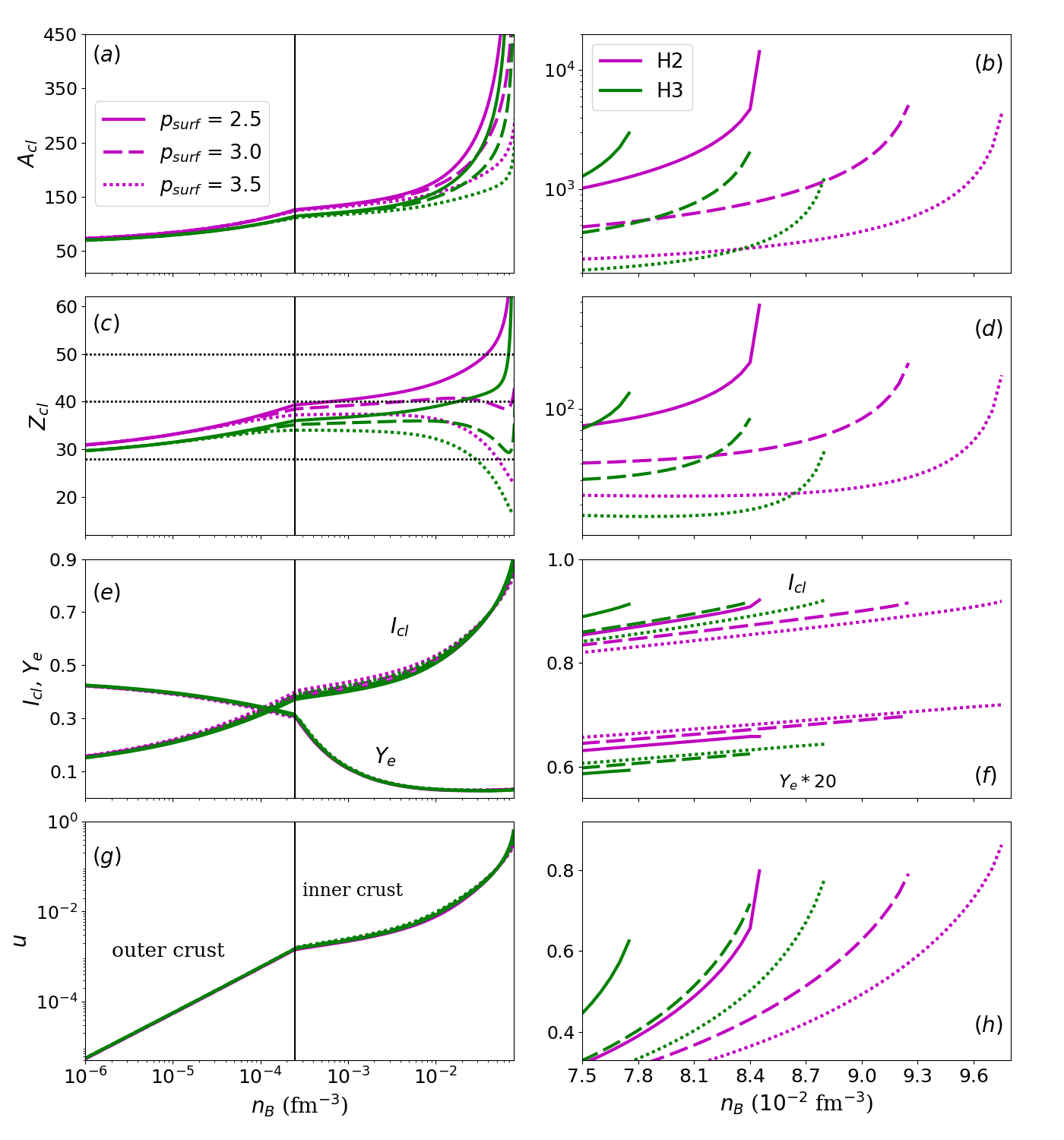}
\caption{Same as Fig. (\ref{fig:crust7H-FS4baremass}). Comparison of tree different values of the surface parameters $p_{\surf}$, within the  H2$_\MMs^{\fs4,\Delta_E}$ and H3$_\MMs^{\fs4,\Delta_E}$ models.}
\label{fig:crustH2-varyp}
\end{figure*}

Note that even if the effect of the FS terms on the crust-core transition density $n_\cc$ is small, it can still be discussed from the right panels in Fig.~\ref{fig:FStermsSLy4}. There is a reduction of $n_\cc$ from FS1 to FS2, consequence of the consistent treatment of the cluster density in the Coulomb and surface terms, which decreases at high density in FS2 while it is fixed to be $n_\sat$ at all densities in FS1. The cluster energy thus gets higher in FS2, compared to FS1, as the density increases and the crust-core transition occurs at a lower density. From FS2 to FS3, the positive curvature term contributing to the energy is compensated in the fit to experimental nuclear masses by a reduction of the surface term for densities given by the experimental data (close to $n_\sat^\emp$). As a consequence FS2 and FS3 are quite similar. However, the surface and curvature terms have a different density dependence in FS2 and FS3. Then as the global density increases in the crust, implying a decrease of the cluster density, the FS term in FS3 becomes lower than the FS term in FS2, and as a consequence $n_\cc$ slightly increases. Finally, the exchange Coulomb term is attractive, opposite to the direct Coulomb term, but being a small contribution, it very slightly pushes up $n_\cc$.

\subsection{Impact of the bulk terms}
\label{sec:NSbulk}

In this section, we analyse the impact of the bulk terms by varying the Hamiltonians and by analyzing the effect of the effective nucleon mass. We contrast the predictions obtained from the Hamiltonians H1-H4, which represent the best models reproducing the experimental nuclear masses, with the Hamiltonians H5 and H7 for which the reproduction of the experimental data are poorer.

The crust predictions by the chiral Hamiltonians H1-H5 and H7 within the FS4 model are displayed in Fig.~\ref{fig:crust7H-FS4baremass} similarly to Fig.~\ref{fig:FStermsSLy4}. We notice that the Hamiltonians H5 and H7 -- already excluded from the finite nuclei analysis -- significantly depart from the predictions by the Hamiltonians H1-H4, which define the light gray band in Fig.~\ref{fig:crust7H-FS4baremass}. They predict lower values for $A_\cl$, $Z_\cl$, $I_\cl$ and slightly larger values for $Y_e$ and $u$. Note also that while the nucleon effective mass prescription plays a role, it is much smaller than the uncertainty in the Hamiltonians. The biggest uncertainties are for the composition ($A_\cl$, $Z_\cl$, $I_\cl$), as seen on top panels. Note that the uncertainty in the value of $Z_\cl$ originating from the Hamiltonian H1-H4 is small in the outer crust (about 2-4), larger at the bottom of the inner crust (about 8-10), and then becomes very large close to the core-crust transition (about 20 or more). The uncertainties from our best models H2-H3 is however smaller. For instance, in the inner crust, these models predict quite stable values for $Z_\cl$ up to about $0.01$~fm$^{-3}$: $34 < Z_\cl < 40$. The vertical gray band in Fig.~\ref{fig:crust7H-FS4baremass}(h) shows the width for $n_\cc$ corresponding to the two best Hamiltonians H2 and H3: $n_\cc = 8.4-9.3$~10$^{-2}$~fm$^{-3}$.

\subsection{Impact of the surface parameter $p_\surf$}
\label{sec:psurf}

In Fig.~\ref{fig:crustH2-varyp} we illustrate the dependence of the EOS on the surface parameter $p_\surf$ for a reduced set of models. We thus consider our two best models H2 and H3 with FS4 and we fix the loss function to be $\Delta_E$. We vary $p_\surf$ from 2.5 to 3.5 as suggested in Refs.~\cite{Carreau2019a} and \cite{Carreau2019b}.

Let us remind that the precise value for $p_\surf$ cannot be determined from finite nuclei, since its influence on the surface energy becomes important for isospin asymmetries which are way beyond the experimental ones. It was first claimed in Ref.~\cite{Carreau2019b} and is similarly illustrated in Fig.~\ref{fig:EsurfFS}(b). Fig.~\ref{fig:crustH2-varyp} shows that $p_\surf$ has however a remarkable impact in the high density region of the inner crust, close to the core-crust transition. It largely controls for instance the slope of $Z_\cl$ close to the core-crust transition: a large value of $p_\surf$ (here 3.5) predicts a decrease of $Z_\cl$ below 30 at high density while a low value (here 2.5) predicts a increase of $Z_\cl$ above 40. The stability of $Z_\cl$ at high density is found only for a very specific value of $p_\surf$.

The impact of $p_\surf$ on the core-crust density $n_\cc$ is large and comparable with the impact of the Hamiltonian as shown in Fig.~\ref{fig:crust7H-FS4baremass}.

In conclusion, the impact of $p_\surf$ on the composition of the crust ($A_\cl$, $Z_\cl$, $I_\cl$) in the high density region is large and quite uncontrolled. More microscopic calculations are required in order to better determine the value of $p_\surf$.

\section{Global NS properties}
\label{sec:tov}

The structure of non-rotating neutron stars is provided by the solution of the spherical hydrostatic equations in general relativity, also named the Tolmann-Oppenheimer-Volkof equations~\cite{Tolman1939,Oppenheimer1939},
\begin{eqnarray}
  \frac{dm(r)}{dr} &=& 4\pi r^2\rho(r),   \label{eq:tov} \\
  \nonumber  \frac{dP(r)}{dr} &=& -\rho(r) c^2\Bigg(1+\frac{P(r)}{\rho(r) c^2} \Bigg)\frac{d\Phi(r)}{dr}, \\
  \nonumber  \frac{d\Phi(r)}{dr} &=& \frac{Gm(r)}{c^2 r^2}\Bigg(1+\frac{4\pi P(r) r^3}{m(r) c^2} \Bigg)\Bigg(1-\frac{2Gm(r)}{r c^2} \Bigg)^{-1},
\end{eqnarray}
where $G$ is the gravitational constant, $c$ the speed of light, $P(r)$ the total pressure, $m(r)$ the enclosed mass, $\rho(r)=\rho_B(r)$ is the total mass-energy density and $\Phi(r)$ the gravitational field. $P$ and $\rho$ have contributions from both the baryons ($P_B$, $\rho_B$) and the leptons ($P_L$, $\rho_L$).

\begin{figure*}[t]
\centering
\includegraphics[scale=0.28]{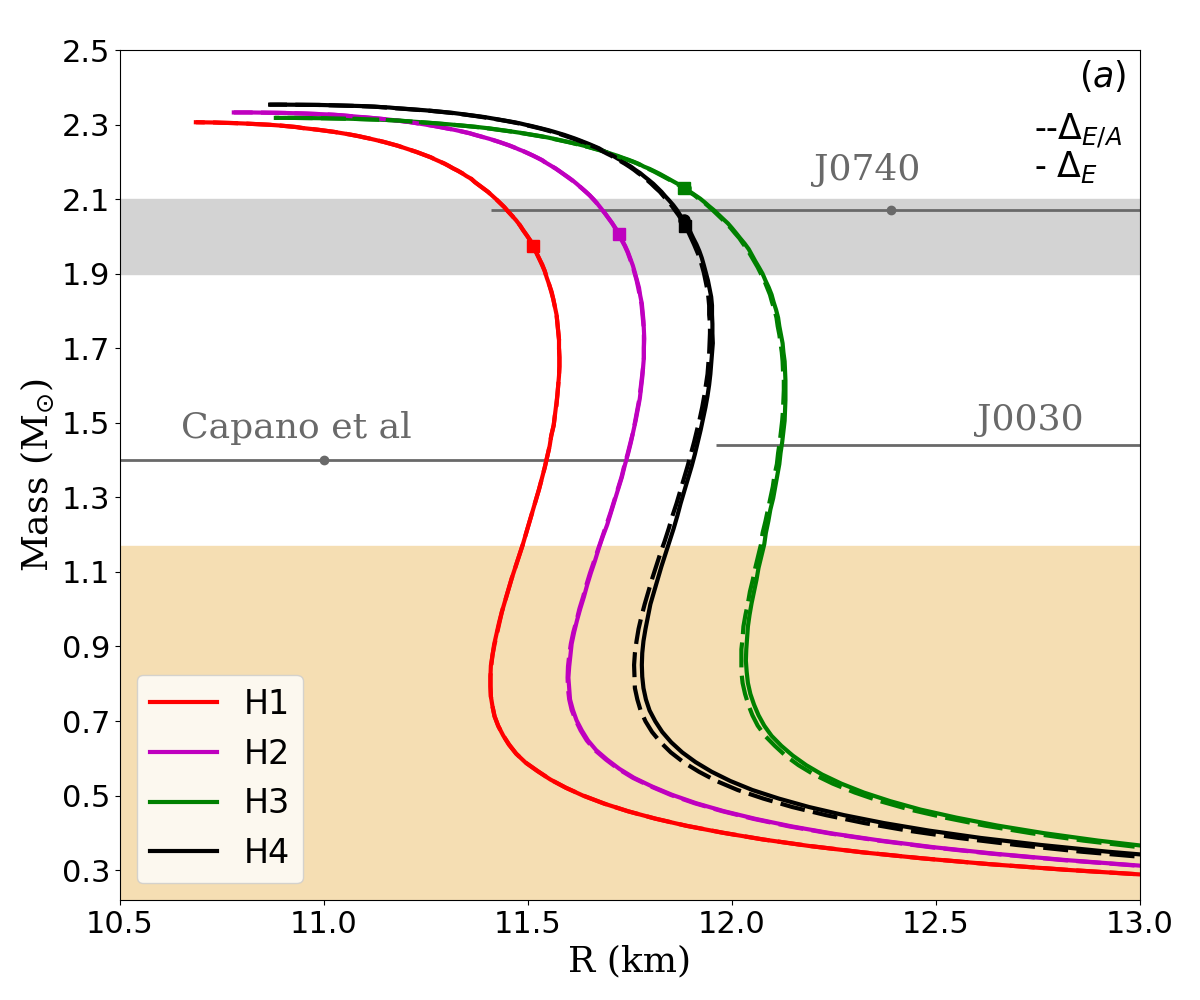}
\includegraphics[scale=0.28]{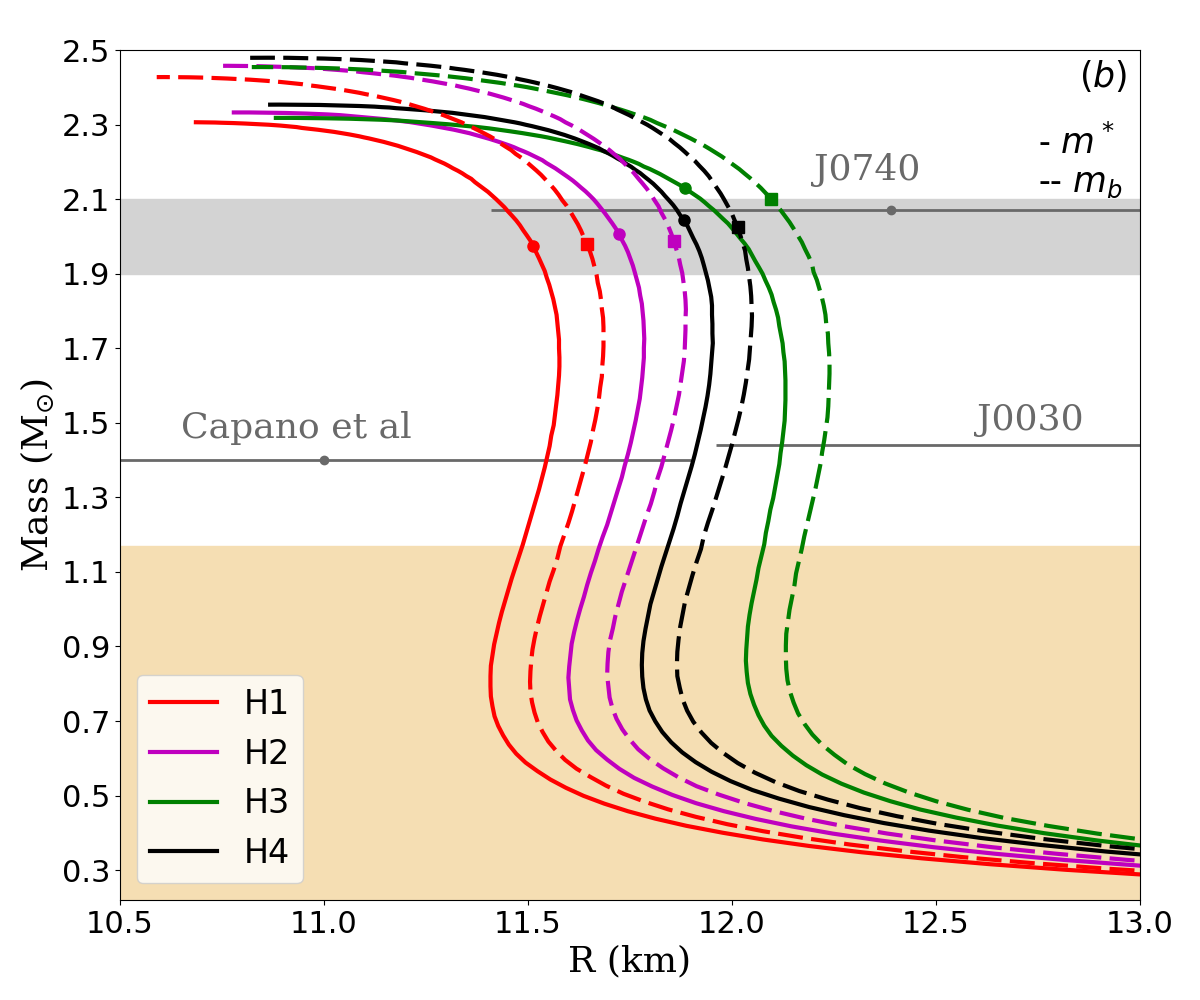}\\
\includegraphics[scale=0.28]{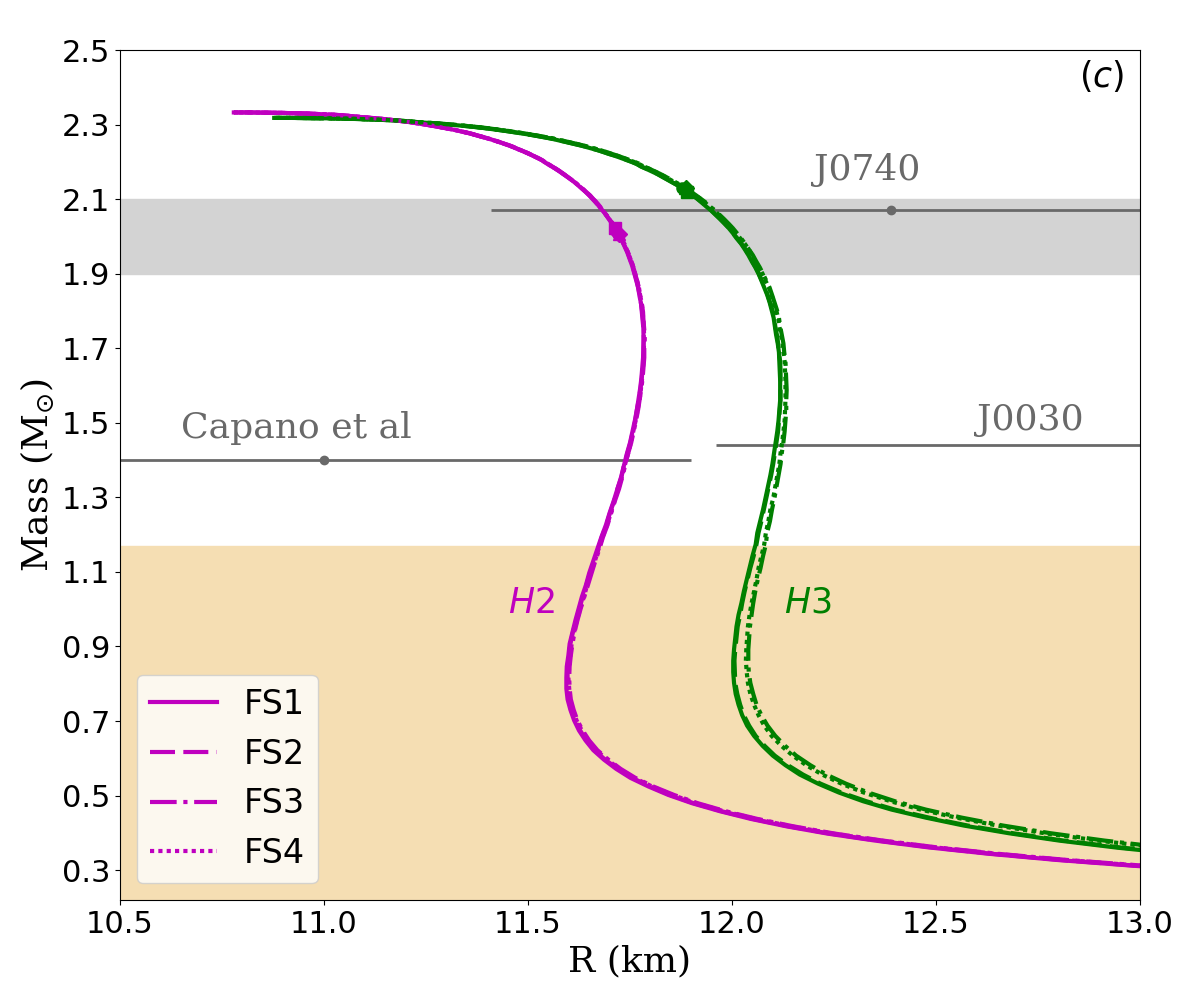}
\includegraphics[scale=0.28]{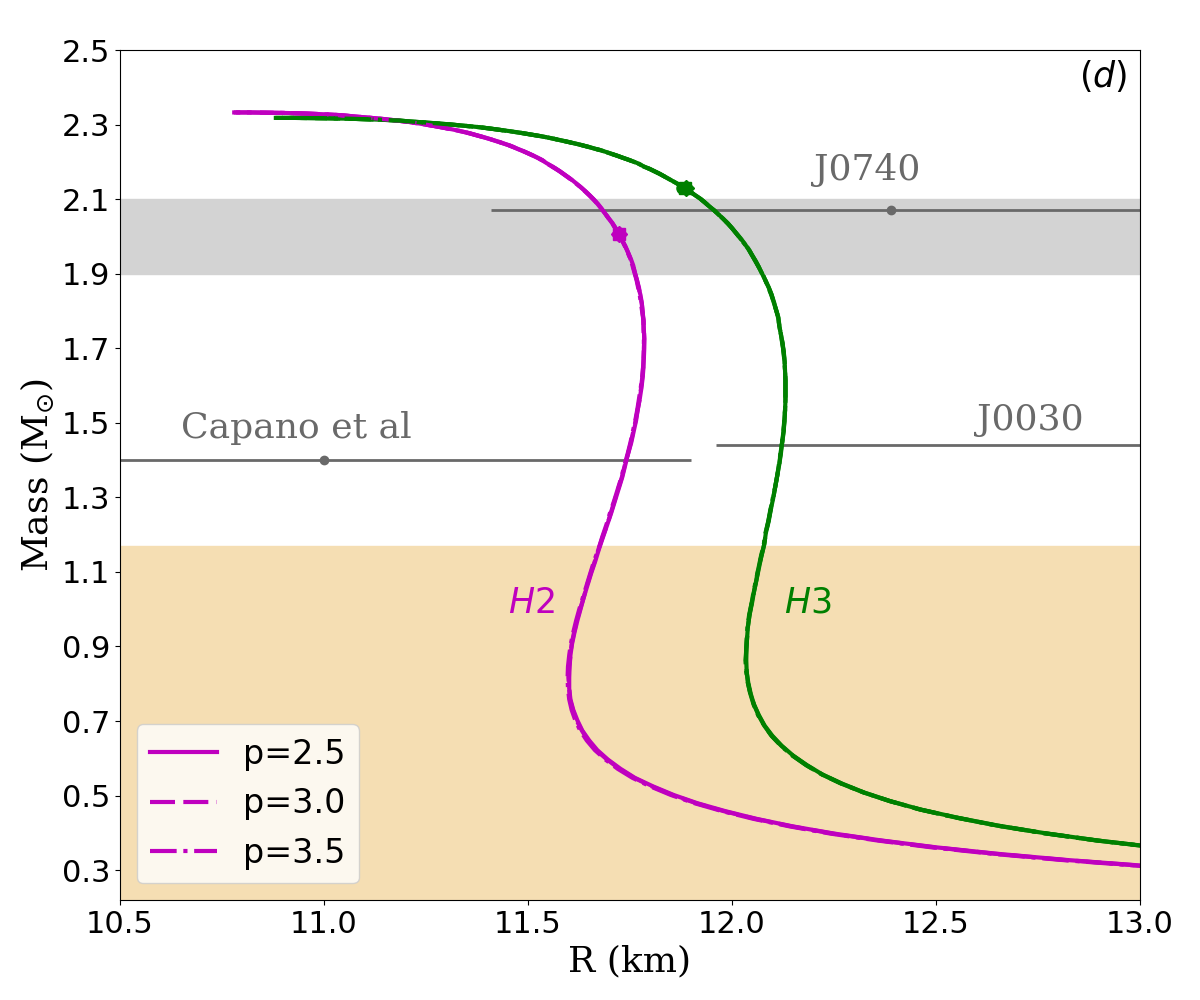}
\caption{Mass and radius relation. Observational constrains from NICER to the pulsar J0030+0451 and J0740+6620, and the constrain from Capano {\it et al.} \cite{Capano2020} combining multi-messengers signals with nuclear physics are shown in error bars. Wheat band shows the lowest mass NS observed with a mass of $1.17 M_{\odot}$. Marks on curves represent the density where causality is broken for a given EOS. Top panels show the results for H1-H4 and (a) Influence of the optimization procedure (b) effect of the inclusion of the nucleon effective mass. Bottom panels show the results for the two Hamiltonians the better reproduce nuclear masses, H2 and H3, and (c) the impact of finite-size terms, (d) the impact of the surface parameter $p_\surf$. See text for more details.}
\label{fig:tov}
\end{figure*}

The four variables ($m$, $\rho$, $P$, $\Phi$) are obtained from the solution of the three TOV equations~(\ref{eq:tov}) and the EOS. In the present calculation, the crust and core EOS are unified, {\it i.e.}, the same nuclear interaction describes crust and core, as seen in the previous sections. The tidal deformability $\Lambda_{GW}$ induced by an external quadrupole field is expressed in terms of the Love number $k_2$ as $\Lambda_{GW}=2k_2/(3C^5)$, where the compactness is $C=GM/(Rc^2)$, and $k_2$ is calculated from the pulsation equation at the surface of NS~\cite{Hinderer2008,Flanagan2008},
\begin{eqnarray}
k_2 &=& \frac{8 C^5}{5}\left( 1-2C\right)^2\left(2-y_R+2C(y_R-1)\right)\nonumber \\
&&\hspace{-1cm}\times\Big(2C(6-3y_R+3C(5y_R-8))\nonumber \\
&&\hspace{-1cm}+4C^3\left(13-11y_R+C(3y_R-2)+2C^2(1+y_R)\right)\nonumber \\
&&\hspace{-1cm}+3(1-2C)^2\left(2-y_R+2C(y_R-1)\right)\ln(1-2C) \Big)^{-1} \!\!\!\! ,
\end{eqnarray}
where $y_R$ is the value of the $y$ function at radius $R$, $y_R=y(r=R)$, and $y(r)$ is the solution of
the following differential equation,
\begin{equation}
r\frac{dy}{dr}+y^2+yF(r)+r^2Q(r)=0\, ,
\end{equation}
with the boundary condition $y(0)=2$ and the functions $F(r)$ and $Q(r)$ defined as,
\begin{eqnarray}
F(r)&=&\frac{1-4\pi r^2 G[\rho(r)-P(r)]/c^4}{1-2M(r)G/(rc^2) } \, ,\\
r^2Q(r) &=& \frac{4\pi r^2 G}{c^4}\left(5\rho(r)+9P(r)+\frac{\partial \rho(r)}{\partial P(r)}[\rho(r)+P(r)]\right)\nonumber \\
&&\hspace{-1.2cm}\times\left(1-2M(r)G/(rc^2)\right)^{-1} - 6\left( 1-2M(r)G/(rc^2) \right)^{-1} \nonumber \\
&&\hspace{-1.2cm}-\frac{4 G^2}{r^2c^8}\left( M(r)c^2+4\pi r^3P(r)\right)^2\left(1-2M(r)G/(rc^2)\right)^{-2} \, .\nonumber \\
\end{eqnarray}

The NS moment of inertia is obtained from the slow rotation approximation~\cite{Hartle1967,Morrison2004} as
\begin{equation}
I = \frac{8\pi}{3}\int_0^R dr r^4 \rho(r)\left(1+\frac{P}{\rho(c)c^2}\right)\frac{\bar{\omega}}{\Omega}e^{\lambda-\Phi} \, ,
\label{eq:moi}
\end{equation}
where $\bar{\omega}$ is the local spin frequency, which represents the general relativistic correction to the asymptotic angular momentum $\Omega$ and $\lambda$ is defined as $\exp(-2\lambda)=1-Gm/(rc^2)$. The crust moment of inertia $I_\crust$ is deduced by considering on the contribution of the crust to the general expression~\eqref{eq:moi}.

As usual, for a given EOS the family of solutions is parameterized by the central density or pressure or enthalpy.
The EOS are characterized by their evolution in the mass-radius
diagram, where both masses and radii of compact stars could in
principle be measured, see also Ref.~\cite{Watts2015}.

We show the mass and radius relations for families of NS in Fig.~\ref{fig:tov}. Horizontal error bars show the constrains from NICER analyzis of the pulsar J0030+0451\cite{MillerNICER2019} and J0740+6620\cite{NICER2021}, and the constrain from Capano {\it et al.}\cite{Capano2020} combining multi-messengers signals with nuclear physics. Gray band shows the maximum mass constrain of 2 M$_{\odot} \pm 0.1$ NS \cite{Demorest2010,Antoniadis2013}. Wheat band marks the lowest NS mass observed of $1.17 M_{\odot}$\cite{Martinez2015}. Fig.~\ref{fig:tov}(a) shows the influence of the  Hamiltonians H1-H4 and the loss function, used in the fit to experimental nuclear masses, on the mass-radius diagram. All curves are calculated with FS4 model and include the effective mass $m^*$. The square-symbols on the curves shows the density above which causality is violated for a given EOS. We note that the loss function has a very small influence for the mass-radius relation, especially in the domain of observed NS (we observe a very small effect for H3 and H4 for masses below $1.1M_{\odot}$). The Hamiltonian H1 predicts the smallest radius (about 11.5km for $1.4M_{\odot}$), while H3 the highest one (about 12.3km for $1.4M_{\odot}$). Note that this prediction is also influenced by the extrapolation of the chiral Hamiltonian at high density, which we do not discuss in this paper.

In Fig. \ref{fig:tov}(b) we show the impact of the nucleon mass prescription (bare versus effective mass), considering FS4 finite-size term and $\Delta_E$ loss function. The impact of the nucleon mass prescription is smaller than the one of the Hamiltonian, but it is the largest one among all other terms discussed here. Note that the use of the bare nucleon mass makes all EOSs a bit stiffer compared to the use of the effective mass, but the overall impact on the radius is not larger than about 100~m (less than 1\% of the radius). The maximum mass is also influenced by the effective mass prescription. We remark that it is increased by $0.15M_{\odot}$ between the effective mass and the bare mass.

In Fig. \ref{fig:tov}(c) we show the mass-radius predictions for the two best Hamiltonians, H2 and H3, for which the different FS approximation are shown, fixing the nucleon effective mass and the $\Delta_E$ loss function. We note that the finite-size terms has essentially no effect on the global mass-radius relation.

Finally, in Fig. \ref{fig:tov}(d) we show the effect of varying the surface parameter $p_\surf$ for H2 and H3, fixing FS4, $m^*$ and the $\Delta_E$ loss function. Although this parameter has shown to be important for the composition and crust-core transition density, for the macroscopic quantities shown Fig. \ref{fig:tov}(d) we see no effect of $p_\surf$.

\begin{figure}[t]
\centering
\includegraphics[scale=0.35]{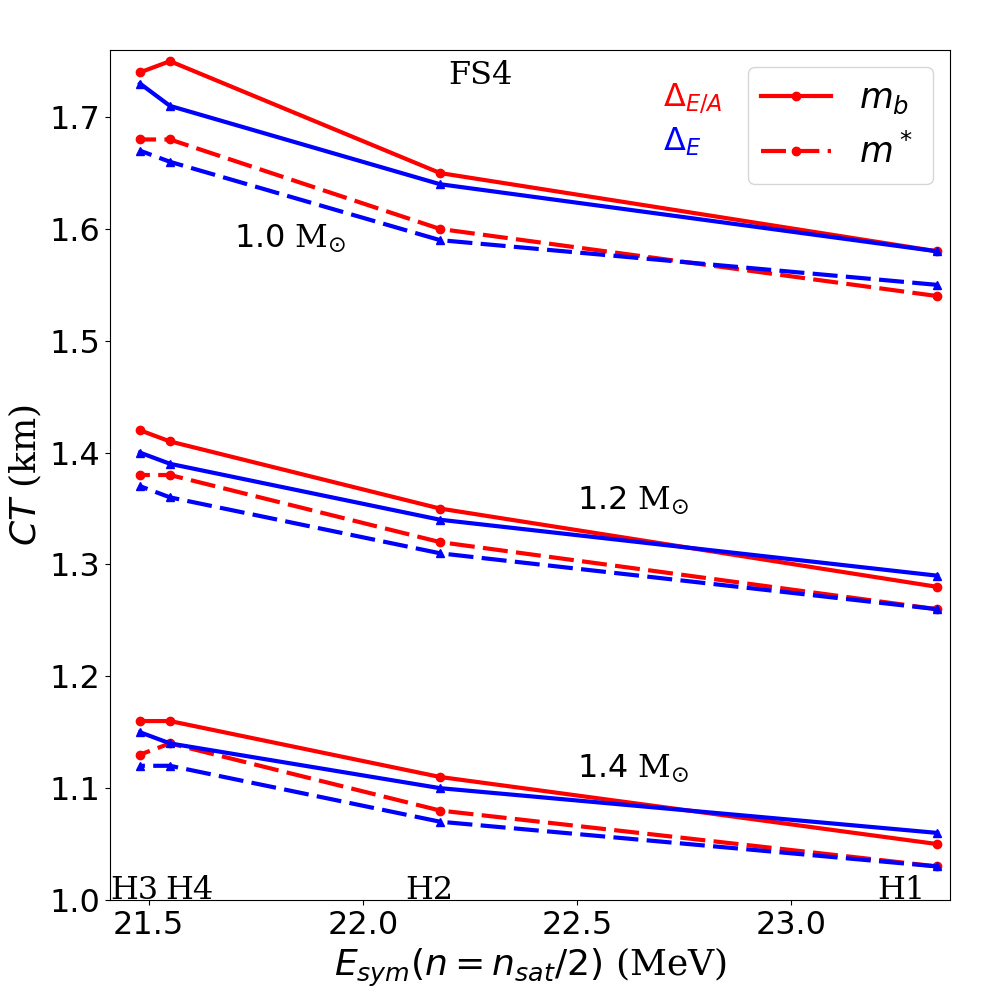}
\caption{Correlation of crust thickness for a 1.0 $M_\odot$, 1.2 $M_\odot$ and 1.4 $M_\odot$ NS and the symmetry energy evaluated at $n_\sat /2$. Dashed lines includes the nucleon effective mass. }
\label{fig:RcrustEsym}
\end{figure}

While the global mass-radius relation is rather insensible to the details of the energetic modeling of the crust, some more specific properties could be. We show for instance in Fig.~\ref{fig:RcrustEsym} the correlation of the crust thickness, $CT$, with the symmetry energy evaluated at half saturation density for three NS masses. We note that the higher $E_\sym (n=n_\sat/2)$ the smaller the crust thickness. By changing the effective mass and the loss function, we also note small effects: the effective mass reduces CT by about 5\% compared to the bare mass, while $\Delta_E$ reduces by about 1\% CT compared to $\Delta_{E/A}$.

We have performed a more systematical study of the uncertainties related to the crust properties which is shown in Table \ref{table:crustPropAv}. We present in this table the values obtained by averaging over the different parameters, as well as the standard deviations reflecting their present uncertainties.

We compare low mass NS ($1.0M_\odot$) with canonical mass NS ($1.4M_\odot$) for the following quantities:
radius, crust thickness, tidal deformability, and crustal moment of inertia. In more details, we have varied the bulk term from H1 to H4, with and without effective mass within the two loss functions, keeping however FS4 fixed. We note a $\approx 3 \%$ uncertainty on the radii. Note that this error could be larger if one assumes a non-unified EOS, as shown in Ref.~\cite{Fortin2016}. For the crust thickness we found an error of $\approx 6 \%$. The size of the crust impacts directly the crust moment of inertia, therefore we see an error of the same order for $I_\crust/I$. Here we bring attention again for the importance of using a unified EOS, since this error could be as high as $10 \%$ \cite{Fortin2016} in case of a not consistent treatment for the NS crust and core. Regarding the tidal deformability of the 1.4 M$_\odot$ NS, the models presented in this work predict $\Lambda_{1.4} = 337 (\pm 56)$, therefore all models are inside the constrain from the GW170817 event \cite{Abbott2017}.

\begin{table}[t]
\centering
\tabcolsep=0.22cm
\def\arraystretch{1.5}
\begin{tabular}{clccc}
\hline\hline
   $R_{1.0}$ (km) &  $CT_{1.0}$ (km) &  $\Lambda_{1.0}$  & (I$_{\crust}$/I)$_{1.0}$  ($\%$)  \\
 11.84 (0.36) &  1.65 (0.10)  &   2040 (375)   &     5.80 (0.42)\\[0.15cm]
  $R_{1.4}$ (km) &  $CT_{1.4}$ (km) &  $\Lambda_{1.4}$  &  (I$_{\crust}$/I)$_{1.4}$ ($\%$) \\
 11.91 (0.34) &  1.10 (0.07)  &   337 (56)   &     3.25 (0.25) \\
\hline\hline
\end{tabular}
\caption{Average value ($\pm$ uncertainty) on neutron star radius, crust thickness (CT), tidal deformability $\Lambda$ and fraction of crust moment of inertia for a 1.0 and 1.4 M$_{\odot}$ NS. Comparison among the four Hamiltonians selected from the finite nuclei study, H1-H4, the different fit prescriptions and the inclusion or not of the nucleon effective mass. Noting that the biggest uncertainty comes from the Hamiltonian choice (H1-H4).}
\label{table:crustPropAv}
\end{table}

\section{Conclusions}
\label{sec:conclusion}

In the present paper we produce a set of compressible liquid-drop models (CLDM) at different orders in the finite-size terms. These CLDM are qualified as unified models for the NS crust and core since the bulk contribution in these two regions are derived from the same model. Here we use the meta-model fitted to a set of Hamiltonians constructed from chiral EFT interactions. Based on these ingredients, we build a set of unified EOS which allows us to explore a number of approximations leading to uncertainties in the crust EOS. We have then investigated the impact of various nuclear approximations and model uncertainties on the prediction for NS crust properties and global quantities. While the present crust model could be extended in the future, we reached some conclusions on the respective impact of various sources of uncertainties which are already interesting for the understanding of NS modeling and of their predictive power.

The confrontation of the CLDM to the experimental nuclear masses presented in this work allowed us to exclude four Hamiltonians: H5, H7, DHS$_{L59}$ and DHS$_{L69}$. We also predicted for the first time the existence of an upper limit for the energy density at saturation $\epsilon_\sat^\mathrm{max}$, induced by the finite-size terms. Based on the CLDM and chiral EFT Hamiltonians, we obtained the following value for this upper limit: $\epsilon_\sat^\mathrm{max}\approx -2.30$~MeV~fm$^{-3}$. 
The good models considered here (H2 and H3) lead to an even more refined value for the energy density at saturation which is $\epsilon_\sat^\best = -2.70(20)$~MeV~fm$^{-3}$.
We have also discussed the effect of including the curvature term in the CLDM: this term leads to a substantial reduction of the residuals between the CLDM and the experimental data, and it also impact the correlation of the surface parameters $\sigma_{\surf,\sat}$ and $\sigma_{\surf,\sym}$ versus the symmetry energy $E_\sym$. In the latter case, we found that the model dependence of this correlation is minimal for our best Hamiltonians H2 and H3. The origin of this effect is still to be understood.

The analysis of NS crust shows that the finite-size terms impact directly the crust composition ($A_\cl$, $Z_\cl$, and $I_\cl$) and very marginally $Y_e$ and the volume fraction $u$. The impact of the finite-size terms on the core-crust transition density $n_\cc$ is also small, while non negligible. $n_\cc$ is however very largely impacted by the Hamiltonian and the surface energy parameter $p_\surf$. We finally make predictions on the EOS properties based on the best models we have (H1-H4 for the wider ones and H2-H3 for the smaller ones). Our results somehow establishes new boundaries based on our present knowledge on nuclear physics ($\ChiEFT$ and experimental nuclear masses) and NS crust modeling (represented here by the different finite-size terms employed in the CLDM).
These boundaries are important for the modeling of NS phenomenology, such as for instance NS cooling or the understanding of QPO in x-ray flares,  oscillation modes in the NS crust as well as spin glitches observed in young NS. In the future, the present approach could be enriched by incorporating the contribution of pairing, neutron skins, different geometries, and of shell effects, and better determination of the parameter $p_\surf$ shall also be explored. 

Finally, the uncertainties on NS macroscopic properties are dominated by the Hamiltonians themselves. 
A realistic estimation of this error requires however a unified description of nuclear matter in the crust and the core of NS, as implemented in this work.

\begin{acknowledgements}
G.G., J.M. and R.S. are supported by CNRS grant PICS-08294 VIPER (Nuclear Physics for Violent Phenomena in the Universe), the CNRS IEA-303083 BEOS project, the CNRS/IN2P3 NewMAC project, and benefit from PHAROS COST Action MP16214. S.R. is supported by Grant No. DE-FG02-00ER41132 from the Department of Energy , and the Grant No. PHY-1430152 (JINA Center for the Evolution of the Elements), and PHY-1630782 (Network for Neutrinos, Nuclear Astrophysics, and Symmetries (N3AS)) from the National Science Foundation. This work is supported by the LABEX Lyon Institute of Origins (ANR-10-LABX-0066) of the \textsl{Universit\'e de Lyon} for its financial support within the program \textsl{Investissements d'Avenir} (ANR-11-IDEX-0007) of the French government operated by the National Research Agency (ANR).
\end{acknowledgements}


\bibliography{ref}
\end{document}


\title{Supplemental material}

\author{G. Grams}
\email{g.grams@ip2i.in2p3.fr}
\affiliation{Univ Lyon, Univ Claude Bernard Lyon 1, CNRS/IN2P3, IP2I Lyon, UMR 5822, F-69622, Villeurbanne, France}

\author{R. Somasundaram}
\affiliation{Univ Lyon, Univ Claude Bernard Lyon 1, CNRS/IN2P3, IP2I Lyon, UMR 5822, F-69622, Villeurbanne, France}
 
\author{J. Margueron}
\affiliation{Univ Lyon, Univ Claude Bernard Lyon 1, CNRS/IN2P3, IP2I Lyon, UMR 5822, F-69622, Villeurbanne, France}

\author{S. Reddy}
\affiliation{Institute for Nuclear Theory, University of Washington, Seattle, WA 98195-1550, USA}

\date{\today}

\begin{abstract}
In this supplemental material we give details about the fit on the nuclear chart. The values of the parameters for each interaction used on the present work are also given. These parameters are used on the CLDM to described the clusterized phase on neutron star crust.
\end{abstract}

\maketitle

\section{Fit to experimental masses}

We confront the compressible liquid drop model (CLDM) against experimental nuclear masses. In this section we give more information about the results of the fit within the four finite size models FS1-4 used in this work.

It was suggested in Refs.~\cite{Steiner2008,Tews2016} to fine tune the FS terms by imposing the CLDM to reproduce the binding energies over the experimental nuclear chart. We proceed similarly in the present study by introducing variational parameters $\mathcal{C}_\coul$, $\mathcal{C}_{\surf, \sat}$, $\mathcal{C}_{\surf, \sym}$, $\mathcal{C}_{\curv, \sat}$, $\mathcal{C}_{\beta}$ correcting the standard values given in Tab. IV of the paper. We have:
\begin{eqnarray}
\sigma_{\surf, \sat} &=& \mathcal{C}_{\surf, \sat} \sigma_{\surf, \sat}^{\std}\, ,\\
\sigma_{\surf, \sym} &=& \mathcal{C}_{\surf, \sym} \sigma_{\surf,\sym}^\std\, , \\ 
\sigma_{\curv, \sat} &=& \mathcal{C}_{\curv, \sat} \sigma_{\curv, \sat}^{\std}\, , \\
\beta_\mathrm{curv} &=& \mathcal{C}_\beta \beta_\mathrm{curv}^{\std} \, .
\end{eqnarray}
If the standard parameters are well chosen, the values for these variational parameters shall remain close to 1.

For each nucleus defined by ($A,Z$) running over the nuclear chart, the variable $n_\cl$ is obtained by imposing mechanical equilibrium, $P_\nuc = n_\cl^2 \partial E_\nuc / \partial n_\cl = 0$. 

The fit strategy consists in optimising the parameters over the nuclear chart considering the experimental energies from the 2016 Atomic Mass Evaluation (AME) \cite{AME2016}, with nuclei in the range: $A = 12 - 295$ and $Z = 6 - 118$.
In practice, we minimize the cost functions $\Delta_{E}$ or $\Delta_{E/A}$, which are defined as
\begin{eqnarray}
\Delta_{E} &=& \left[ \frac{1}{N} \sum_{i=1}^{N} ( E_\ex^i - E_\nuc^i )^2 \right]^{1/2}\, , \\
\Delta_{E/A} &=& \left[ \frac{1}{N} \sum_{i=1}^{N} \left( \frac{E_\ex^i}{A} - \frac{E_\nuc^i}{A} \right)^2 \right]^{1/2} \, ,
\end{eqnarray}
where $N=3375$ is the number of considered nuclei from the experimental nuclear chart, $E_\ex^i$ the experimental values, and $E_\nuc^i$ the energy from the CLDM model.

For FS1 and FS2, the minimization is performed by varying the three scaling factors $\mathcal{C}_\coul$, $\mathcal{C}_{\surf, \sat}$, $\mathcal{C}_{\surf, \sym}$,  while for FS3 and FS4, we vary in addition the scaling factors $\mathcal{C}_{\curv, \sat}$, and $\mathcal{C}_\beta$. We use the well-known Gauss-Newton algorithm, which converges fast.
The parameter $p_\surf=3$ is fixed since finite nuclei do not constrain this parameter~\cite{Carreau2019a}.

The nuclear interaction, e.g. Skyrme, Gogny or RMF, are usually determined from the optimisation of the cost function $\Delta_{E}$, while the CLDM is sometimes fit by optimising the cost function $\Delta_{E/A}$~\cite{Carreau2019a}.
Since there is an ambiguity in the cost function to employ, we consider in our analysis the two methods and we estimate the impact of this uncertainty in the fit for the properties of the NS crust.

\begin{table*}[t]
\centering
\tabcolsep=0.3cm
\def\arraystretch{1.5}
\begin{tabular}{cccccccccc}
\hline\hline
Model & FS & Minimization & $\mathcal{C}_\coul$ & $\mathcal{C}_{\surf, \sat}$ & $\mathcal{C}_{\surf, \sym}$ & $\mathcal{C}_{\curv, \sat}$ &  $\mathcal{C}_{\beta}$ & $\Delta_E$  &  $\Delta_{E/A}$ \\
 &  &  & & & & & & (MeV) & (MeV) \\
\hline\hline
SLy4$_\mm$ & FS1 & $\Delta_{E}$&  0.9470 &  1.0354 &  1.0172 &  0.0 &  0.0 &   3.32 &   0.054\\
 & FS1 & $\Delta_{E/A}$&  0.9510 &  1.0231 &  0.9011 &  0.0 &  0.0 &   4.02 &   0.048\\
\hline
SLy4$_\mm$ & FS2 & $\Delta_{E}$&  0.9408 &  1.0514 &  1.1322 &  0.0 &  0.0 &   3.33 &   0.048\\
 & FS2 & $\Delta_{E/A}$ &  0.9336 &  1.0568 &  1.0841 &  0.0 &  0.0 &   3.71 &   0.045\\
\hline
SLy4$_\mm$ & FS3 & $\Delta_{E}$&  0.9614 &  0.9649 &  1.1731 &  1.5171 &  0.9943 &   2.76 &   0.047\\
 & FS3 & $\Delta_{E/A}$&  0.9439 &  1.0235 &  1.1390 &  0.6857 &  0.9309 &   3.12 &   0.038\\
\hline
SLy4$_\mm$ & FS4 & $\Delta_{E}$&  0.9684 &  1.0322 &  1.2543 &  1.1340 &  0.9812 &   2.72 &   0.046\\
 & FS4 & $\Delta_{E/A}$ &  0.9524 &  1.0841 &  1.2016 &  0.3686 &  0.8497 &   3.08 &   0.038\\
\hline\hline
\end{tabular}
\caption{Optimization of the scaling factor governing the CLDM binding energy over the nuclear chart, where SLy4$_\mm$ interaction for the bulk term is used with different finite size terms ($\fs i$) and minimizing either $\Delta_{E}$ or $\Delta_{E/A}$.}
\label{table:nucleifit:SLy4}
\end{table*}

The resulting scaling factors obtained for the different FS models (FS1 to FS4) as well as by minimizing either $\Delta_{E}$ or $\Delta_{E/A}$ are given in Tab.~\ref{table:nucleifit:SLy4}. In this table, the homogeneous matter contribution is fixed to be the one given by SLy4$_\mm$. 
The Skyrme SLy4 interaction~\cite{Chabanat1997} has been employed in several predictions for the NS crust~\cite{DouchinHaensel2000,DouchinHaensel2001,Vinas2017,Carreau2019a}, since it was fit to both nuclei and low-density neutron matter as predicted by the microscopic calculations of Wiringa {\it et al.} \cite{Wiringa1988}.

For the models FS1 and FS2, the scaling factors $\mathcal{C}_\coul$ and $\mathcal{C}_{\surf, \sat}$ are weakly impacted (less than 1\%) by the choice of the cost function, at variance with the asymmetry factor $\mathcal{C}_{\surf, \sym}$, which is largely impacted by both the choice of the cost function (about 5-10\%) and the dependence of the Coulomb and surface term on $n_\cl$ (about 10-20\%).
For FS1 and FS2, minimizing $\Delta_{E/A}$ instead of $\Delta_E$ over the nuclear chart reduces $\mathcal{C}_{\surf, \sym}$ by 5-15\%.
The cost functions $\Delta_E$ and $\Delta_{E/A}$ do not treat identically the experimental data over the mass table. Since the total energy $E$ scales with the mass number, the cost function $\Delta_E$ tends to prefer the large mass region for the fit, while the cost function $\Delta_{E/A}$ treats in a more equal way light and heavy nuclei. 
The surface asymmetry parameter $\mathcal{C}_{\surf, \sym}$ reflects the isospin constraints induced by the two prescriptions, since light and intermediate mass nuclei explore larger asymmetries than the heavy ones. 

Note that minimising $\Delta_E$ induces a slightly larger minimum value for FS2 than for FS1, while minimizing $\Delta_{E/A}$ induces a reduction of the cost function from FS1 to FS2, as expected. It is not an anomaly, but it reflects the complexity of defining a single global function, such as $\Delta_E$ or $\Delta_{E/A}$, to measure the quality of the fit.

The curvature term introduced in FS3 impacts the parameters of the surface term (by 5-10\%), almost not the Coulomb factor (less than 2\%) and the curvature factor are largely impacted by the choice of the cost function for the fit (more than 50\% correction for $\mathcal{C}_{\curv, \sat}$). There is also a clear gain in the quality functions $\Delta_E$ and $\Delta_{E/A}$ coming from the introducing of the curvature term.

The introduction of the Coulomb exchange term in FS4 does not strongly reduced the quality functions, but its influence on the scaling factors themselves is noticeable: about 5-10\% for $\mathcal{C}_{\surf, \sat}$ and $\mathcal{C}_{\surf, \sym}$ and 30\% for $\mathcal{C}_{\curv, \sat}$.
The scaling factors $\mathcal{C}_\coul$ and $\mathcal{C}_\beta$ are less impacted.
The r.m.s. of the residual that we obtain for SLy4$_\mm$, $\Delta_E=2.72-2.76$~MeV, is comparable to the one found in Ref.~\cite{Steiner2008}: $\Delta_E=2.6$~MeV. The latter is slightly better since the associated model contains one more parameter (associated to neutron skin), as discussed on the paper.

In Table \ref{table:nucleifit:hamiltonians} we show the resulting scaling factors obtained for the different Hamiltonians (H1-H7, except H6) with and without effective mass, optimizing either the total energy ($\Delta_{E}$) or or the energy per baryon ($\Delta_{E/A}$) keeping FS4 as the only FS model. The resulting cost functions $\Delta_{E}$ and $\Delta_{E/A}$ are shown in the two last columns. Note that H2 and H3 obtain the lowest cost function among the Hamiltonians, as discussed on the paper.

\begin{table*}[t]
\centering
\tabcolsep=0.3cm
\def\arraystretch{1.5}
\begin{tabular}{lccccccccc}
\hline\hline
FS & Model & Fit & $\mathcal{C}_\coul$ & $\mathcal{C}_{\surf, \sat}$ & $\mathcal{C}_{\surf, \sym}$ & $\mathcal{C}_{\curv, \sat}$ &  $\mathcal{C}_{\beta}$ & $\Delta_E$  &  $\Delta_{E/A}$ \\
 & & & & & & & & (MeV) & (MeV) \\
\hline\hline
FS4 & H1$_\MMb^{\fs4}$& $E$ &  0.9666 &  1.5722 &  1.6081 &  1.1458 & -0.0199 &   3.038 &   0.049\\
 & H1$_\MMb^{\fs4}$& $E/A$ &  0.9667 &  1.5690 &  1.5727 &  0.5787 & -0.6264 &   3.270 &   0.041\\
 & H1$_\MMs^{\fs4}$& $E$&  0.9655 &  1.5825 &  1.3083 &  0.9418 & -0.2584 &   3.183 &   0.051\\
 & H1$_\MMs^{\fs4}$& $E/A$&  0.9665 &  1.5760 &  1.3368 &  0.5546 & -0.7674 &   3.673 &   0.043\\
\hline
FS4 & H2$_\MMb^{\fs4}$ & $E$  &  0.9273 &  1.0274 &  1.3520 &  1.2651 &  1.1231 &   2.744 &   0.051\\
 & H2$_\MMb^{\fs4}$& $E/A$  &  0.9089 &  1.0935 &  1.2701 &  0.3367 &  1.2838 &   3.348 &   0.040\\
 & H2$_\MMs^{\fs4}$& $E$  &  0.9273 &  1.0374 &  1.1239 &  1.0534 &  1.1386 &   2.707 &   0.045\\
 & H2$_\MMs^{\fs4}$& $E/A$ &  0.9100 &  1.1001 &  1.0919 &  0.3297 &  1.1694 &   3.062 &   0.038\\
\hline
FS4 & H3$_\MMb^{\fs4}$& $E$ &  0.9127 &  0.8437 &  1.2905 &  1.3275 &  1.4656 &   2.879 &   0.060\\
 & H3$_\MMb^{\fs4}$& $E/A$  &  0.8873 &  0.9336 &  1.2000 &  0.2379 &  2.9389 &   4.062 &   0.044\\
 & H3$_\MMs^{\fs4}$& $E$  &  0.9126 &  0.8476 &  1.0957 &  1.1271 &  1.5575 &   2.771 &   0.054\\
 & H3$_\MMs^{\fs4}$& $E/A$ &  0.8881 &  0.9352 &  1.0474 &  0.2358 &  2.8464 &   3.625 &   0.041\\
\hline
FS4 & H4$_\MMb^{\fs4}$& $E$ &  0.9027 &  0.7135 &  1.1510 &  1.3535 &  1.6914 &   3.044 &   0.068\\
 & H4$_\MMb^{\fs4}$& $E/A$ &  0.8724 &  0.8189 &  1.0635 &  0.1624 &  5.3893 &   4.610 &   0.048\\
 & H4$_\MMs^{\fs4}$& $E$  &  0.9030 &  0.7164 &  0.9681 &  1.1444 &  1.8303 &   2.888 &   0.061\\
 & H4$_\MMs^{\fs4}$& $E/A$ &  0.8736 &  0.8198 &  0.9196 &  0.1652 &  5.1532 &   4.099 &   0.044\\
\hline
FS4 & H5$_\MMb^{\fs4}$& $E$ &  0.8650 &  0.3266 &  0.7577 &  1.3943 &  2.3827 &   4.146 &   0.102\\
 & H5$_\MMb^{\fs4}$& $E/A$&  0.8393 &  0.4143 &  0.7768 &  0.4956 &  4.2853 &   5.537 &   0.075\\
 & H5$_\MMs^{\fs4}$& $E$ &  0.8676 &  0.3257 &  0.6471 &  1.2160 &  2.6133 &   3.863 &   0.095\\
 & H5$_\MMs^{\fs4}$& $E/A$&  0.8426 &  0.4086 &  0.6714 &  0.4797 &  4.4399 &   5.274 &   0.071\\
\hline
FS4 & H7$_\MMb^{\fs4}$& $E$ &  0.8587 &  0.1131 &  0.3466 &  1.0903 &  3.2150 &   5.356 &   0.126\\
 & H7$_\MMb^{\fs4}$& $E/A$   &  0.8097 &  0.2509 &  0.4308 & -0.3798 & -4.1410 &   9.139 &   0.079\\
 & H7$_\MMs^{\fs4}$& $E$  &   0.8636 &  0.1012 &  0.2620 &  0.9411 &  3.6488 &   4.951 &   0.119\\
 & H7$_\MMs^{\fs4}$& $E/A$  &  0.8129 &  0.2474 &  0.3671 & -0.3370 & -4.7318 &   8.434 &   0.074\\
 \hline
FS4 & DHS$_{L59}$ $_\MMb^{\fs4}$& $E$ &  0.8530 &  0.3703 &  1.0383 &  1.6870 &  2.0874 &   4.414 &   0.109\\
 & DHS$_{L59}$ $_\MMb^{\fs4}$& $E/A$  &  0.8163 &  0.4941 &  1.0186 &  0.1950 &  8.8712 &   6.720 &   0.073\\
 \hline
FS4 & DHS$_{L69}$ $_\MMb^{\fs4}$& $E$ &  0.8714 &  0.5669 &  1.5717 &  1.9719 &  1.6477 &   4.099 &   0.102\\
 & DHS$_{L69}$ $_\MMb^{\fs4}$& $E/A$   &  0.8430 &  0.6636 &  1.4769 &  0.5582 &  2.9592 &   6.251 &   0.074\\
\hline\hline
\end{tabular}
\caption{Sames as Table \ref{table:nucleifit:SLy4} but for the eight Hamiltonians with and without effective mass, keeping FS4 fixed. }
\label{table:nucleifit:hamiltonians}
\end{table*}


\bibliographystyle{unsrt}
\bibliography{ref}